\documentclass[pdflatex,sn-nature]{sn-jnl}

\usepackage{graphicx}%
\usepackage{multirow}%
\usepackage{amsmath,amssymb,amsfonts}%
\usepackage{amsthm}%
\usepackage{mathrsfs}%
\usepackage[title]{appendix}%
\usepackage{xcolor}%
\usepackage{textcomp}%
\usepackage{manyfoot}%
\usepackage{booktabs}%
\usepackage{algorithm}%
\usepackage{algorithmicx}%
\usepackage{algpseudocode}%
\usepackage{listings}%
\usepackage{aas_macros} 

\theoremstyle{thmstyleone}%
\theoremstyle{thmstyletwo}%

\theoremstyle{thmstylethree}%

\newcommand{\nastro}{Nat. Astron.} 
\newcommand{\sci}{Science} 
\newcommand{\sciadv}{Sci. Adv.} 

\newcommand{\ion}[2]{#1\;\textsc{#2}}
\newcommand{\agn}{J2245$+$3743} 
\newcommand{\fullagn}{J224554.84$+$374326.5}
\newcommand{\Lya}{Ly$\alpha$}
\newcommand{\Ha}{H$\alpha$}
\newcommand{\Hb}{H$\beta$}
\newcommand{\civ}{\ion{C}{IV}}
\newcommand{\ciii}{\ion{C}{III}]}

\newcommand{\mgii}{\ion{Mg}{II}}
\newcommand{\oiii}{[\ion{O}{III}]}

\newcommand{\as}[1]{#1$^{\prime \prime}$}

\begin{document}





\title[An extremely luminous AGN flare]{An Extremely Luminous Flare Recorded from a Supermassive Black Hole}


\author*[1]{\fnm{Matthew J.} \sur{Graham}}\email{mjg@caltech.edu}

\author[2,3,4,5]{\fnm{Barry} \sur{McKernan}}

\author[2,3,4,5]{\fnm{K. E. Saavik} \sur{Ford}}

\author[6]{\fnm{Daniel} \sur{Stern}}

\author[5,7]{\fnm{Matteo} \sur{Cantiello}}

\author[1]{\fnm{Andrew J.} \sur{Drake}}

\author[1]{\fnm{Yuanze} \sur{Ding}}

\author[1]{\fnm{Mansi} \sur{Kasliwal}}

\author[8]{\fnm{Mike} \sur{Koss}}

\author[9,10]{\fnm{Raffaella} \sur{Margutti}}

\author[1]{\fnm{Sam} \sur{Rose}}

\author[1]{\fnm{Jean} \sur{Somalwar}}

\author[11]{\fnm{Phil} \sur{Wiseman}}

\author[1]{\fnm{S. G.} \sur{Djorgovski}}

\author[12]{\fnm{Patrik M.} \sur{Veres}}

\author[13]{\fnm{Eric C.} \sur{Bellm}}

\author[14]{\fnm{Tracy X.} \sur{Chen}}

\author[14]{\fnm {Steven L.} \sur{Groom}}

\author[1]{\fnm {Shrinivas R.} \sur{Kulkarni}}

\author[1]{\fnm {Ashish} \sur{Mahabal}}

\affil*[1]{\orgdiv{Division of Physics, Maths and Astronomy}, \orgname{California Institute of Technology}, \orgaddress{\street{1200 E California Blvd}, \city{Pasadena}, \state{CA} \postcode{91125}, \country{USA}}}

\affil[2]{\orgdiv{Department of Science}, \orgname{CUNY Borough of Manhattan Community College}, \orgaddress{\street{199 Chambers Street}, \city{New York}, \state{NY} \postcode{10007}, \country{USA}}}

\affil[3]{\orgdiv{Department of Astrophysics}, \orgname{American Museum of Natural History}, \orgaddress{\street{Central Park West}, \city{New York}, \state{NY} \postcode{10007}, \country{USA}}}

\affil[4]{\orgdiv{Physics Program}, \orgname{CUNY Graduate Center}, \orgaddress{\street{365 5th Avenue}, \city{New York}, \state{NY} \postcode{10016}, \country{USA}}}

\affil[5]{\orgdiv{Center for Computational Astrophysics}, \orgname{Flatiron Institute}, \orgaddress{\street{162 5th Avenue}, \city{New York}, \state{NY} \postcode{10016}, \country{USA}}}

\affil[6]{\orgdiv{Jet Propulsion Laboratory}, \orgname{California Institute of Technology}, \orgaddress{\street{4800 Oak Grove Drive}, \city{Pasadena}, \state{CA} \postcode{91109}, \country{USA}}}

\affil[7]{\orgdiv{Department of Astrophysical Sciences}, \orgname{Princeton University}, \city{Princeton}, \state{NJ} \postcode{08544}, \country{USA}}

\affil[8]{\orgname{Eureka Scientific}, \orgaddress{\street{2452 Delmer Street Suite 100}, \city{Oakland}, \state{CA} \postcode{94602-3017}, \country{USA}}}

\affil[9]{\orgdiv{Department of Astronomy}, \orgname{University of California}, \city{Berkeley}, \state{CA} \postcode{94720}, \country{USA}}

\affil[10]{\orgdiv{Department of Physics}, \orgname{University of California}, \orgaddress{\street{366 Physics North MC 7300}, \city{Berkeley}, \state{CA} \postcode{94720}, \country{USA}}}

\affil[11]{\orgdiv{School of Physics and Astronomy}, \orgname{University of Southampton}, \orgaddress{\street{University Road}, \city{Southampton}, \state{Hants} \postcode{SO17 1BJ}, \country{UK}}}

\affil[12]{\orgdiv{Astronomical Institute (AIRUB)}, \orgname{Ruhr University Bochum}, \orgaddress{\street{Universitätsstraße 150}, \city{Bochum}, \postcode{44801}, \country{Germany}}}

\affil[13]{\orgdiv{DIRAC Institute}, \orgname{Department of Astronomy, University of Washington}, \orgaddress{\street{3910 15th Avenue NE}, \city{Seattle}, \state{WA}, \postcode{98195}, \country{USA}}}

\affil[14]{\orgdiv{IPAC}, \orgname{California Institute of Technology}, \orgaddress{\street{1200 E California Blvd}, \city{Pasadena}, \state{CA}, \postcode{91125}, \country{USA}}}


\abstract{%
Since their discovery more than 60 years ago, accreting supermassive black holes in active galactic nuclei (AGN) were recognized as highly variable sources, requiring an extremely compact, dynamic environment.  Their variability traces to multiple phenomena, including changing accretion rates, temperature changes, foreground absorbers, and structural changes to the accretion disk. Spurred by a new generation of time-domain surveys, the extremes of black hole variability are now being probed. We report the discovery of an extreme flare by the AGN J224554.84+374326.5, which brightened by more than a factor of 40 in 2018. The source has slowly faded since then.  The total emitted UV/optical energy to date is $\sim10^{54}$~erg, i.e., the complete conversion of approximately one solar mass into electromagnetic radiation. This flare is 30 times more powerful than the previous most powerful AGN transient. Very few physical events in the Universe can liberate this much electromagnetic energy. We discuss potential mechanisms, including the tidal disruption of a high mass $(>30\, M_\odot)$ star, gravitational lensing of an AGN flare or supernova, or a supermassive (pair instability) supernova in the accretion disk of an AGN. We favor the tidal disruption of a massive star in a prograde orbit in an AGN disk.
}




\maketitle

\section*{Main Text}\label{sec:MainText}

Supermassive black holes (SMBHs) are host to a wide variety of variable phenomena. Inactive, or quiescent SMBHs can produce tidal disruption events (TDEs) as a star comes too close to an SMBH and is torn apart by the black hole's tidal forces \citep[for a recent review, see][]{Gezari21}.  Active galactic nuclei (AGN), which are SMBHs with accretion disks, demonstrate stochastic variability produced by both coherent and flaring activity \cite[e.g.,][]{Komossa24}. Based on observable quantities such as timescale, energetics, color evolution, and spectral evolution, AGN flares are currently classified into several broad, likely overlapping phenomenological groups, including changing-look AGN \cite[CL-AGN; e.g.][]{LaMassa15, MacLeod19}, changing-state AGN \cite[CS-AGN; e.g.,][]{Stern18, Graham20, Ricci23}, periodic AGN \cite[e.g.,][]{Graham15}, TDEs \cite[e.g.,][]{Masterson22}, quasi-periodic eruptors \cite[QPEs; e.g.,][]{Miniutti19}, quasi-periodic oscillators \cite[QPOs; e.g.,][]{Middleton09}, ambiguous nuclear transients \cite[ANTs; e.g.,][]{Wiseman24}, and extreme nuclear transients \cite[ENTs;][]{Hinkle24}.  The physical mechanisms behind these events are still unclear, but the nature and location of the activity strongly suggests at least some are caused by interactions between the AGN accretion disk and a star, most likely from the nuclear star cluster gravitationally bound to the SMBH \cite[e.g.,][]{Graham20, Starfall22}. Other potential physical mechanisms include gravitational lensing, a supernova in the nuclear region (possibly in the AGN accretion disk), or disk instabilities.

The discovery of SMBH-related transients has been enabled by a number of large time domain surveys over the past decade \cite[e.g.,][]{Drake09, Hodgkin13, Shappee14, Tonry18}. The occurrence rates are low and the events can last for months to years, so they require wide sky coverage with long baselines. The Catalina Real-time Transient Survey \cite[CRTS;][]{Drake09} and Zwicky Transient Facility \cite[ZTF;][]{Bellm19, Graham19} together provide 20 years coverage over 70\% of the sky to a depth of $\gtrsim 20^{th}$ magnitude in optical bandpasses and have supported a number of systematic studies of rare AGN phenomena \cite[e.g.,][]{Graham15, Graham17, Graham23}. Recent detections of very luminous transients $(M < -23)$ associated with AGN (i.e., ANTs and ENTs) that share characteristics with TDEs, superluminous supernovae, and the more extreme end of AGN variability have motivated a search for such sources in the CRTS+ZTF data set.

\fullagn\, (hereafter \agn) is an AGN at $z = 2.554$ associated with the most luminous AGN flare reported to date. A brightening by almost four magnitudes (corresponding to a flux increase by a factor of 40) was detected in early 2018 by CRTS and ZTF (see Fig.~\ref{fig:lightcurve}). The associated host is consistent with an AGN based on its mid-infrared (mid-IR) colors \cite{Stern12}, though no pre-flare spectrum of the source is available. The flare slowly dimmed over the subsequent six years (observer frame; $\sim 650$ days in the restframe) but the flux still remains two magnitudes above the pre-flare level. A sequence of spectra over the past six years show the gradual development of broadened emission lines---e.g., \Lya, \civ, \ciii, \mgii, \Hb, \oiii, and \Ha---consistent with an AGN.  The spectra also show absorption systems indicative of both a weak ($\lesssim 1000$ km s$^{-1}$) outflow and multiple intervening systems along the line-of-sight. Using \civ, \mgii, and \Ha\ emission lines, we estimate a single-epoch SMBH virial mass of $10^{8.5 \pm 0.3}\, M_\odot$.

WISE detects the flare at mid-IR wavelengths with no statistically significant time lag relative to the optical flare, and the flare was undetected at X-ray and radio wavelengths. There is also no detection of associated neutrino emission, as has been seen in very energetic TDEs with mid-IR emission \cite{Stein21, vanVelzen24}. The total radiated energy associated with the transient to date is $\sim 10^{54}$~erg (isotropic equivalent).

The characteristic restframe rise and decay timescales, $\tau_{\rm rise} \simeq 40$ days and $\tau_{\rm decay} \simeq 140$ days, of \agn\ are comparable to other ENTs \cite{Hinkle24}, but the peak luminosity, $L_{pk} \sim 4 \times 10^{46}$ erg s$^{-1}$, and total energy, $\sim 10^{54}$~erg, are one to two orders of magnitude larger than any previously published AGN flare (see Fig.~\ref{fig:absmagtime}). In fact, there are few phenomena easily capable of producing so much energy. If the \agn\ flare is a TDE, then it might involve a star with a mass much larger than $20~M_{\odot}$, assuming 10\% for the efficiency rate of converting accreted mass to emitted energy. While a supernova is ruled out on energetics, it could be a superluminous supernova embedded in the AGN accretion disk, i.e., most of the energy is provided by the kinetic energy of dense surrounding material \cite{Grishin21}. If the flare is a gravitationally lensed event then the actual total energy involved would be less, i.e., the observed flux is the true flux magnified by some factor, but the event would still be extreme for typical observed lensing magnifications. High resolution imaging shows no indication of multiple components and simulations show that the statistical likelihood of lensing is very low. The most likely explanation is then a massive TDE in an AGN disk.

Almost all TDEs detected to date have been in quiescent, i.e., non-AGN, low redshift galaxies (although there is certainly a degree of selection bias).  Only five TDEs are currently reported at $z > 1$ \cite{Gu24}. At $z = 2.554$, \agn\ would be both the most energetic and the highest redshift TDE detected to date (we note that since submission of this paper a photometric TDE candidate has been reported at $z = 5.02^{+1.32}_{-1.11}$ \cite{Karmen25}). This suggests that there is certainly a population of less energetic TDEs associated with AGNs at $z > 1$ which could aid in probing the characteristics and evolution of SMBH environments at high redshift. In order to achieve the observed total energy released, the mass of the disrupted star is estimated to be $\gtrsim 30 M_\odot$, presumed drawn from a population of massive stars in the AGN accretion disk. Further examples of luminous TDEs in AGN will probe the proposed top-heavy stellar mass function of stars in AGN disks. Timely high energy observations and high angular resolution imaging will also help to confirm this explanation in future cases.

\section*{Discussion}\label{sec:discussion}

The brightest nuclear transients are equally attributed to a number of physical mechanisms, partly due to a lack of any clear characteristic separation between current models. \cite{Hinkle24} show this degeneracy for peak absolute magnitude and characteristic (half decay) timescale (see Fig.~\ref{fig:absmagtime}). A full multidimensional treatment considering photometric, spectroscopic, and multiwavelength properties may allow a clearer distinction between models but this is outside the scope of this current work. 

Several explanations might produce the energy observed in \agn, but we can rule many out from  straightforward arguments. For example, we disfavour strong beaming or a remarkably strong embedded GRB \citep[e.g.,][]{Perna21} since much of the released energy should appear across a wide range of wavelengths (including X-rays). However, there are only upper limits to X-ray emission from \agn\ during the flare. Furthermore, diffusion of emission should dominate, which means the peak luminosity should be $\ll 10^{45}$ erg due to isotropization of beamed afterglow emission, which is inconsistent with observations of \agn. Another means of  generating a flare of this magnitude is SMBH self-lensing of an AGN flare. The probability of yielding a SMBH micro-lensed flare that yields no change in lightcurve or color as both the position of the AGN flare and the SMBH move relative to each other on a timescale of $\sim {\rm yr}$ (in the restframe) is vanishingly small so we disfavour this explanation.

We also disfavour an individual nuclear supernova as an explanation for \agn\ for several reasons. First, a total energy of $10^{54}$ erg exceeds the maximum total energy (including kinetic energy of the explosion) allowable from pair instability supernovae (PISN) by an order of magnitude \citep{Woosley02,Woosley21}. On the far side of the pair-instability gap ($M_{\ast} \gtrsim 240M_{\odot}$), core collapse is expected to resume, but strong winds are expected to carry away most of the envelope, so the entire massive stellar core is believed to collapse into a single, massive BH, with no (or no significant) electromagnetic emission \citep{Renzo24}. A more detailed consideration of each explanation is presented in the Supplementary Information. We next consider \agn\ in terms of the most plausible remaining explanations, a TDE. 

\subsection*{The case for a tidal disruption event}
TDEs are the most common transient phenomena associated with quiescent SMBHs. TDEs occur when a star of mass $M_{\ast}$ and radius $R_{\ast}$ passes quickly within a SMBH tidal radius ($R_{\rm T}$) which is larger than the SMBH physical radius \citep{Rees88}, i.e.,
\begin{equation}
R_{\rm T} \approx R_{\ast}\left( \frac{M_{\rm SMBH}}{M_{\ast}}\right)^{1/3} > 2 \frac{GM_{\rm SMBH}}{c^{2}}.
\end{equation}
Though most TDEs identified to date are in quiescent galaxies, 
TDEs must also occur in AGN \citep{Chan19}. The rate of AGN-TDEs is expected to be higher than that for quiescent nuclei early in the  AGN lifetime as the disk acts to alter existing stellar orbits and rapidly fills the orbital loss-cone \citep[e.g.][]{Starfall22,Wang24,Ryu24,Prasad24}. Classical TDEs involve stars with masses $\sim 0.5 - 2 M_\odot$ around SMBHs with $M_{\rm SMBH} \leq 10^8\, M_\odot$, and such events release $\leq 10^{51}$ erg of energy. Candidate AGN-TDEs are significantly more energetic than classical TDEs since the disrupted star can interact strongly with the AGN disk \citep{Chan19,Ryu24}. The star-disk interaction can trap otherwise escaping ejecta and drive high velocity interactions between the TDE accretion stream and the AGN disk  \citep{Starfall22,Ryu24,Wang24}. 

The three most luminous nuclear transients reported to date \citep[i.e., ENTs;][]{Hinkle24} (see Fig.~\ref{fig:tde} and Supplementary Table 1) show long-lived ($>150$ day) flares that are far more energetic -- $0.5 - 2.5 \times 10^{53}$ erg -- than classical TDEs and are thought to be due to AGN-TDEs of intermediate mass ($\sim3 M_\odot < M < 10 M_\odot$) stars. We note that the brightest, AT2021lwx (also known as Scary Barbie) \cite{Subrayan23}, was initially attributed to a 14 $M_\odot$ TDE using the MOSFIT tool \cite{Guillochon18}. We find that MOSFIT consistently reports extreme TDEs to be $\sim$14 $M_\odot$ stars disrupted by $\sim$ 10$^8 M_\odot$ SMBHs, suggestive of a fitting artifact. Other TDE-fitting tools are also not suitable for ENTs or TDEs in AGN. The ANT-comparable transient AT2023vto \cite{Kumar24} (also on Fig.~\ref{fig:absmagtime}) is attributed to the tidal disruption of a $\sim 9.1 M_\odot$ star by a quiescent $\sim 10^7 M_\odot$ black hole and is spectroscopically similar to He-class TDEs (i.e., TDEs with prominent helium lines), though AT2023vto is significantly brighter than the brightest known classical TDE. Intriguingly, TDEs of massive stars may also be  associated with the advent of compact symmetric radio AGN \citep{Readhead24}.

\agn\ is more energetic by a factor 5-10 than all previously reported ENTS. Therefore, if powered by an AGN-TDE, this suggests $M_{\ast} \gg 10M_{\odot}$. Simple energetic scaling suggests the disruption of a star in the mass range [30,100]$M_{\odot}$. However, if we assume all the energy originates from an accretion event of luminosity $L_{\rm acc}$ over a characteristic timescale $t_{\rm char}$, then $L_{\rm acc} t_{\rm char} \sim \eta \langle \dot{M} \rangle c^{2} t_{\rm char} \sim 10^{54}\,{\rm erg}$. This yields an average accretion rate $\langle \dot{M} \rangle \sim 5.5 M_{\odot}\, {\rm yr}^{-1}\, (\eta/0.1)^{-1} (t_{\rm char}/\rm{yr}) \sim \dot{M}_{\rm Edd}(M_{\rm SMBH}/3 \times 10^{8}M_{\odot})$. So, assuming that up to half the debris could in principle escape as in standard TDEs, $M_{\ast} \gtrsim 11M_{\odot}$.

It has long been believed that massive stars can form in AGN disks \citep{GoodmanTan04,Nayakshin07}. Some support for this idea comes from our own Galactic nucleus where a top heavy initial mass function ($M^{-0.5}$) is observed for disky stars (presumably formed in an AGN episode) \citep{Bartko10}. More recently, simulations have been carried out for models of stellar evolution of stars embedded in semi-realistic AGN disk models 
\citep{Cantiello21,Jermyn22,Chen25}, suggesting that stars embedded in dense AGN disks might evolve up to several hundred solar masses over $\mathcal{O}$(Myr) AGN lifetimes \citep{DaviesLin20,Fabj24}. So, massive stars in AGN are not unexpected. Interestingly, a TDE of a  massive star ($\sim 30\, M_{\odot}$) implies a scattering event off a more massive object (e.g., a more massive star, e.g. $\mathcal{O}(100\, M_{\odot})$, an intermediate mass black hole (IMBH; $\gtrsim 100\, M_{\odot}$), or a binary black hole in the IMBH mass range from further out in the disk. Thus, TDEs of massive stars test the dynamics among the embedded population in AGN disks \citep[e.g.][]{Wang24,Prasad24}.

The Hills mass of an SMBH above which an electromagnetic signature is not possible given an in-plane TDE is given by eqn. 64 in \citep{Mummery24} as 
\begin{equation}
 M_{\rm Hills} = \left[\frac{5 R_\ast^3 c^6}{\eta G^3 M_\ast}\right]^{1/2}\frac{1}{[1 + \sqrt{(1 - a)}]^3}   
\end{equation}
or $M_{\rm Hills} \sim 8.8\times 10^7M_{\odot}$ in the traditional limit where the spin $a=0$, the stellar radius $R_\ast=R_\odot$, the stellar mass $M_\ast = M_\odot$, and $\eta \sim \mathcal{O}(1)$ is a factor which incorporates a parameterization of stellar structure (including rotation). Note that this is smaller than the value $M_{\rm SMBH} \sim 3  \times 10^8M_{\odot}$ estimated for this AGN.
However, if we assume $R_{\ast} \sim 10\, R_\odot, M_\ast \sim 35\, M_\odot$ and $a=0$, then the above result scales by
$M_{\rm Hills} \sim 8.8 \times 10^7\, M_\odot \times (10^3/35)^{1/2} \sim 4.7\times 10^8\, M_\odot$, or $>M_{\rm SMBH}$. So the Hills mass for a massive star (assuming $\eta \sim \mathcal{O}(1)$) is nearly a factor of two larger than our estimate of $M_{\rm SMBH}$ in this AGN and we are not forced to assume a spinning SMBH ($a>0$). If instead we drop the mass of the putative disrupted star to $M_{\ast} \sim 10M_\odot$ (our approximate lower limit, see above) with $R_\ast \sim 5R_\odot$ and $a \sim 0$, then $M_{\rm Hills}\sim 8.8 \times 10^7M_{\odot} \times (5^3/10)^{1/2} \sim 3.1\times 10^8M_\odot$, or $M_{\rm Hills} \sim M_{\rm SMBH}$. So, while we are not strictly forced to assume the SMBH is Kerr ($a>0$), there is support for $a>0$ if $M_{\rm SMBH} \sim 10^{8.5}\, M_\odot$ is an underestimate. If the SMBH is spinning, this increases $M_{\rm Hills}$ by up to a factor $\mathcal{O}(8)$ for  maximal prograde SMBH spin ($a \sim 1$). 

A highly spinning SMBH would also allow for the possibility of jet formation associated with this event and some associated level of beaming early on, at least, where we have limited observational constraints (see the discussion of beaming below). This would have the additional effect of reducing the required energy associated with the TDE by factors of several. However, the fact that no X-rays are observed $\mathcal{O}(500)$ days post event suggests, at best, a short-lived or failed jet. \cite{Bloom11} show a mildly relativistic jet associated with a TDE, releasing $\sim 2 \times 10^{53}$\, erg in X-rays over the first 50 days. Such an event, beamed in our direction, could have been detected. In addition, a jet would be expected to sweep up circumnuclear material and decelerate over the long timescale of this flare, which would result in a change in the slope of the decay (especially for an off-axis jet), which is not seen.

We have discussed how a population of over-massive stars in the AGN nuclear environment is not unexpected. The question then arises, how do massive stars end up with a pericenter sufficient to be disrupted? The excess of AGN-TDEs over random classical TDEs arises from two main sources: (i) a scattering event that fills the loss-cone with the scattered star and (ii) a star on a retrograde orbit flips to prograde as it  rapidly decays towards the SMBH. 

In (i) the scattering event can occur in the inner or outer disk. An inner disk scattering seems less likely as it is harder to alter the velocity vector of an embedded orbiting massive star. Only a close (chaotic) dynamical encounter with a significantly more massive binary \citep{Starfall22,Prasad24}, or a very close encounter between a massive star and a retrograde orbiter with very high relative velocity \citep{Wang24} could fill the loss cone. A scattering event in the outer disk is more likely to deflect even a relatively massive star from an eccentric orbit into the loss cone \citep{Prasad24} and we regard this as more likely. A massive star near apoapse in the outer disk travels relatively slowly. A close encounter with a comparable mass object (or even a less massive object at higher relative velocity) is far more likely to perturb the velocity vector of the massive star into the loss cone than if the star were in the inner disk. As the star approaches the SMBH, it perturbs the disk gas and interacts immediately and violently with the inner disk \citep{Ryu24}, leading to a prompt, energetic flare. The form of \agn\ so far is suggestive of a prograde AGN-TDE, since we see no evidence that the flare has decayed to below the initial continuum level, as would be expected for a retrograde AGN-TDE  \citep{Starfall22}.

In (ii) an initially retrograde star that is flipped by disk torquing to prograde is likely to reach the tidal disruption radius before it flips to an in-plane prograde orbit. Thus, in (ii) we generally expect an inclined TDE with respect to the AGN disk. In this case, the interaction between the inclined TDE disk and the AGN disk should lead to shocks and collisions at small disk radii, generating significant X-ray luminosity \citep{Chan19}, which we do not observe. 

The long duration of the flare requires a strong ongoing interaction between the TDE debris and the disk, which implies that the orbit of the star is at most modestly inclined relative to the AGN disk, if not actually in-plane. In a classical TDE around a quiescent SMBH, half of the mass of the star escapes to infinity but the AGN disk prevents that when a TDE occurs in an AGN. There is also no evidence of dimming of the AGN immediately pre-flare so the flare is not powered by the circularization of self-colliding returning bound debris. This again rules out a strong inclination with respect to the disk with the bulk of the TDE mini-disk being established in the first year or so after the disruption event. 

A classical TDE around a quiescent SMBH should show a $L(t) \propto t^{-5/3}$ decline \cite{Phinney89}. A TDE involving a massive star would be more likely to exhibit a shallower decline, \cite[e.g., $t^{-2/3}$;][]{Guillochon13}. This is because the disruption of a massive star leads to a higher peak fallback rate, which in turn drives a more powerful and sustained high Eddington accretion phase. This accretion is expected to launch substantial, optically thick outflows that then dominate the observed emission through processing. This leads to a shallower light curve decay governed by the outflow's expansion and cooling rather than the raw fallback rate.


\subsection*{Expected rates}
A lower bound to the rate of transients like \agn\ can be estimated from:
\[ R = \sum_{i=1}^N \frac{1}{t_{\mathrm{survey, i}}/(1 + z_i)} \frac{1}{V_{\mathrm{max},i}}f_{\mathrm{loss}} \]

\noindent 
where $t_{\rm survey,i}$ is the time span of the survey detecting the $i^{th}$ transient, corrected into the restframe by $(1 + z_i)$ and $V_{\rm max,i}$ is the maximum comoving volume in which that transient would be detected by the survey. The completeness factor $f_{\mathrm{loss}}$ is the fraction of the sky observed by the survey. $N$ is the total number of observed transients (here we set $N=1$). A transient source normally at the nominal survey detection limit for ZTF of $g = 20.5$ needs to brighten above $g = 20$ to generate a $5 \sigma$ alert. We take $t_{\mathrm{survey}}$ = 6.75 yr and $f_{\mathrm{loss}}$ = 0.73, which gives a rate of $R = 3 \times 10^{-4}$ Gpc$^{-3}$ yr$^{-1}$. For comparison, \cite{Hinkle24} estimate the ENT rate as $R_{\mathrm{ENT}} = 1.0_{-0.6}^{+1.0} \times 10^{-3}$ Gpc$^{-3}$ yr$^{-1}$ and a TDE rate at $z \sim 1$ of $R_{\mathrm{TDE}} \simeq 20 - 80$ Gpc$^{-3}$ yr$^{-1}$.

The stellar mass function proposed for AGN disk stars is $dN / dm \propto m^{-0.45 \pm 0.3}$ \cite{Bartko10}, compared to a standard Salpeter/Kroupa initial mass function of $dN / dm \propto m^{-2.15 \pm 0.3}$. If we associate classical TDEs (in quiescent systems) with stars with a mass range of $0.5 - 2\, M_{\odot}$, ENTs with stars with a mass range of $3 - 10\, M_{\odot}$ and \agn-like TDEs with stars in the mass range $30 - 100\, M_{\odot}$ and assume a standard mass-luminosity relation for main sequence stars then the Salpeter IMF gives relative rates between TDEs and J2245-like events that are too small by a factor of roughly 100. For the AGN stellar mass function, the relative rates are too large by a a factor of 10 - 20. However, a $M_\odot$ star embedded in an AGN disk can become massive ($M_\ast \sim 10^2 - 10^3 M_\odot)$ within $\leq$ Myr and then collapse violently. If we assume a lifetime for such an embedded star of $\sim 1$ Myr then the relative rates broadly match those determined above. 

\section*{Conclusions}\label{sec13}

\agn\ is a singular event: a flare in an existing AGN with the energy equivalent of a solar mass, for which there are few viable physical mechanisms. The absence of any detected high energy flux argues against strong beaming or an embedded GRB. Any form of gravitational lensing is statistically unlikely and high resolution imaging shows no indication of lensed components or a lensing galaxy. The light curve of \agn\ also shows no sign of multiple contributions. An embedded supernova is possible but the most likely progenitor needs to be extremely massive which again makes a supernova statistically unlikely. This leaves a massive TDE in an AGN as the most plausible explanation. The disrupted star is also likely to be on a prograde orbit either in-plane or modestly inclined to the disk. The inferred rate of such events, albeit from a single instance, also broadly agrees with the expected rate for stars in AGN disks assuming a top heavy mass function. 

At $z = 2.554$, \agn\ is the highest redshift candidate TDE to date. This strongly suggests that there is a population of $z > 1$ TDEs in AGNs waiting to be uncovered, and which have the potential to be valuable probes of the central SMBH environment and the accretion disk. They could also provide insight into driving mechanisms for AGN variability, particularly propagating thermal fronts associated with changing-look quasars \cite{Stern18, Wang24}. The challenge is distinguishing a TDE signal from other potential events in the AGN disk and generic AGN variability. 

We are certain that forthcoming deep, wide-field time domain surveys with the Rubin Observatory and the Roman Space Telescope will detect further examples and allow for more dedicated followup of the most massive TDEs. We also suspect that some of the brightest AGN flares already detected over the past twenty years may be similar such instances. A dedicated reappraisal may allow for further insight through their late time (many years post-flare) evolution.

\section*{Methods}\label{sec:methods}

\subsection*{Photometric data}

The \agn\ flare was first detected by the Zwicky Transient Facility \cite[ZTF;][]{Bellm19, Graham19, Masci19, Dekany20} on UT 2018 April 5, leading to the broadcast of alert packet ZTF18acehgts announcing this event to the community. The flare was subsequently identified by the Catalina Real-time Transient Survey \cite[CRTS;][]{Drake09} as a transient on UT 2018 June 6 (MLS180606:224555+374326), although the first detection of the source in a bright state within the CRTS MLS alert system was on UT 2018 May 25. CRTS reported it as virtual observatory event (VOEvent) with a candidate detection and suggested classification, which was posted to the Rochester Bright Supernovae site (\url{www.rochesterastronomy.org/sn2018/index.html}). A few weeks later, on UT 2018 June 28, it was detected by the Asteroid Terrestrial-impact Last Alert System \cite[ATLAS;][]{Tonry18}, which reported it to the Transient Name Service (\url{wis-tns.org}) on UT 2018 July 1 where it was assigned the designation AT~2018cym \cite{Tonry18}. 

Prior photometry of the source starting in 2010 was obtained from both the Catalina Sky Survey (CSS) component of CRTS and the Panoramic Survey Telescope and Rapid Response System \cite[Pan-STARRS;][]{Chambers16}. These show a source with a median magnitude of $r = 21.9$. ZTF data is calibrated to the photometric system defined by the first Pan-STARRS data release \cite{Masci19} and so direct comparison of PS1 and ZTF photometry is possible. To enable similar comparison between CSS and ZTF photometry, forced PSF photometry \cite{Masci23} was carried out at the source position. As CSS data is unfiltered, the light curve was calibrated and transformed to PS1 $r$-band magnitudes using fits to the PS1 $g$-, $r$-, and $i$-band magnitudes \cite{Magnier20}. Individual photometric measurements were corrected for non-linearity and spatial variations in the camera response and atmospheric transmission following the Zubercal process [{\bf Drake et al., in prep.}]. This provides significant improvement in span, quality, and depth compared to public CSS aperture photometry available from CRTS.

We assume that the pre-flare magnitude of the source represents the host contribution and remove that to estimate the flare flux alone (see Supplementary Figure 1). The magnitudes were corrected for Galactic extinction using a standard dust map \citep{Schlegel98} to get line-of-sight $E(B-V)$ values assuming $A_V = 3.1\, E(B-V)$. We also assume a flat $K$-correction of $K(z) = -2.5 \log_{10}(1 + z)$, consistent with the transient flare literature \citep[e.g.,][for ENTs]{Hinkle24}. Assuming a flat $\Lambda$CDM cosmology with $H_0 = 69.6$ km s$^{-1}$ Mpc$^{-1}$ and $\Omega_m = 0.29$, we obtain a peak absolute $r$-band magnitude for \agn\ of $M_r = -27.2$ at MJD = 58295.


We determine the color evolution of \agn\ (see Supplementary Figure 2) by modeling the bandpass light curves as Gaussian processes with Ornstein-Uhlenbeck kernels, which are appropriate for AGNs \cite{Kelly09}, and calculating the differences. Errors on the colors are given by adding in quadrature the standard deviations from the Gaussian process fits. \agn\ has a strong red color with a median post-peak value of $g - r = 0.56$ (and $c - o = 0.28$). The flare shows no discernible color evolution thus far, i.e., 650 days post-peak in the restframe.

\subsection*{Spectroscopic data}

An initial optical spectrum was taken with the Double Beam Spectrograph \cite[DBSP;][]{Oke82} on the Hale \as{200} (P200) telescope at Palomar Observatory on UT 2018 November 2. Observing conditions were poor that night and the lack of any apparent strong emission features in the spectrum led to a speculative association with the nearest radio source NVSS~J224553+374253, 35$^{\prime \prime}$ away, and a suggested blazar classification for the transient. Subsequent longslit optical spectra were taken in 2023 and 2024 (see Table~\ref{tab:spectra} and Supplementary Figure 3) with DBSP on the Hale telescope and the Low Resolution Imaging Spectrometer \cite[LRIS;][]{Oke95} on the Keck-I 10-m telescope at the W. M. Keck Observatory. These data were reduced using standard techniques within IRAF. 

A near-IR spectrum was obtained on UT 2024 September 11 with the Near-Infrared Echelle Spectrograph \cite[NIRES;][]{Wilson04} on the Keck-II telescope. The data, which have spectral resolution $R = \lambda / \Delta \lambda = 2700$, were reduced using a custom version of the IDL-based reduction package {\tt Spextool} \cite{Cushing04} modified for use with NIRES and using {\tt xtellcor} \cite{Vacca03} to correct for telluric features in the spectrum using an A0 standard star observed close in airmass and time to the target. Supplementary Figure 3 shows a composite spectrum combining the LRIS spectrum from UT 2024 July 4 and the NIRES spectrum scaled to equal flux at the \mgii\ $\lambda 2798$ line.

The spectra taken since 2023 ($>500$ days post-peak in the restframe) show broad emission lines consistent with an AGN at $z = 2.554$, where the redshift is determined from the [\ion{O}{III}] $\lambda 5007$ line in the NIRES spectrum. Emission line ratios place the source very clearly in the AGN part of the Baldwin, Phillips \& Terlevich (BPT) diagram \cite{Baldwin81}. The AGN spectral energy distribution template of \cite{Temple21} provides a good fit to each individual spectrum with no indication of significant deviation except for the NIRES spectrum where the \Ha\ and \Hb\ emission lines are weaker than expected.  The full-width half-maxima (FWHMs) of the broad UV lines -- \Lya, \civ, \ciii, and \mgii\ -- are between 3000 and 4000 km s$^{-1}$ but the optical Balmer lines -- \Ha\ and \Hb\ -- are narrower, with values of $\sim 1000$ km s$^{-1}$ or less. This is consistent with some narrow-line Seyfert 1 galaxies in the local universe \cite[e.g.,][]{RodriguezPascal97}.
 
All optical spectra (see Supplementary Figure 4) also show narrow ($\sim 500 - 1200\, {\rm km}\, {\rm s}^{-1}$) P-Cygni absorption features at the position of the broad \Lya\ and \civ\ emission lines, implying some form of outflow. The strengths of these emission lines increase with time but the P-Cygni features remain constant. This type of spectral behavior is seen in a subset of changing-state AGN which are ``turning on" \cite[e.g., see][]{Graham20}. 
\Hb\ is also weak relative to \Ha.  This apparent suppression may be due to absorption of \Hb\ by a dense outflow, as seen in some TDEs, e.g., PS1-10jh and AT2018zr.

The optical forbidden oxygen lines, [\ion{O}{ii}] $\lambda 3726,3728$, and \oiii\ $\lambda 4959,5007$, and [\ion{O}{i}] $\lambda 6300$, are clearly visible in the NIRES spectrum (see Supplementary Figure 5). They all show an asymmetric profile with a blueshifted wing, which is commonly treated as an indicator of outflows related to the central engine in AGN \cite{Schmidt21}. The blueside extension of the wing component can be quantified by a double Gaussian decomposition of the line profile, defining the radial velocity difference to be 
$\Delta v - \frac{1}{2} \mathrm {width}_{\mathrm{[OIII]\, wing}}$, where $\Delta v$ is the velocity difference between the centroids of the two components. We measure a value of $-1168$ km s$^{-1}$ for \oiii\, which is consistent with an SMBH mass of $10^{7.6 \pm 0.5} M_\odot$ \cite{Schmidt21}. 

The [\ion{O}{ii}] emission line doublet can be strongly excited by star formation within the host galaxies of AGN, excited in low density extended emission line regions (EELRs) far from the nucleus. A discriminating factor is the presence of [\ion{Ne}{V}] $\lambda3426$, as seen in the NIRES spectrum, which is not strongly excited by star formation \cite{Maddox18}. As this is a highly ionized neon line, it also suggests that the EELR region is strongly ionized by the active nucleus in \agn. This is similar to AT2019aalc, a candidate repeating TDE in an AGN \cite{MilanVeres24}.

\subsection*{Imaging data}
\agn\, is detected in PS1 images as a compact source with some faint indication of extended morphology to the south-east. Deeper imaging in $B$ and $i$ was obtained with LRIS on UT 2024 October 1 with a total of $6 \times 300$s exposures taken in a dithered sequence (see Supplementary Figure 6). The data were reduced using the IDL {\tt lpipe} pipeline \cite{Perley19} with an astrometric correction determined by {\tt astrometry.net} \cite{Lang10} applied.  

The flare is coincident with the center of the host (to within 5 kpc projected distance at the redshift of the source), confirming its likely identification as a nuclear transient. An extension is evident in the LRIS $i$-band image (and faintly in the $B$-band). The residual from subtracting the estimated transient flux from the $i$-band image using an effective point spread function constructed from the $i$-band image with {\tt photutils} \cite{Bradley24} and matched to the transient flux shows no secondary source brighter than $i = 23.1$. 

Optical integral field unit (IFU) observations were conducted with the Keck Cosmic Web Imager (KCWI) on Keck-II on UT 2024 December 4. The instrument was aligned in the north-south direction (position angle PA = 0°). The total effective exposure time was 2400s for the blue grating and 1800s for the red grating. Each 1200s blue exposure is paired with $3\times300$s red channel exposures in order to mitigate the high cosmic ray (CR) contamination on the red CCD. Observing conditions were excellent, with seeing ranging from \as{0.4} to \as{0.6} as measured by the Canada-France-Hawaii Telescope  Differential Image Motion Monitor. KCWI was configured with the small slicer and the BL and RL gratings, centered at wavelengths of 4500\AA\ and 7250\AA\, respectively, covering spectral ranges of 3420–5600\AA\ and 5600–9030\AA\ with the 5600\AA\ dichroic. 

In this setup, the slitlets are \as{0.35} wide in the east-west direction, with a pixel scale of \as{0.147} per pixel in the north-south direction and an \as{8.4} × \as{20.4} field of view. The average spectral resolution is $R \simeq 3600$ for both gratings. Twilight flat fielding was conducted before taking any science images. Flux calibration was performed using the standard star Feige 110, observed immediately prior to the science target. We combined three standard star exposures, after refitting the data reduction pipeline (DRP) inverse sensitivity curve. The telluric corrections were performed separately with the Pypelt-based \citep{Prochaska20} data reduction tool KCWISkywizard. Data reduction for the blue channel was primarily completed with the standard KCWI DRP. We created an additional sky mask from the intermediate 2D images to remove astronomical sources from the sky model. A medium filtering background subtraction was conducted before the Differential Atmospheric Refraction (DAR) correction so as to correct the amplifier differences between the left and right part of the CCD, utilizing the custom tool KcwiKit \citep{KcwiKit}. The red channel reduction was conducted by a customized pipeline, which read in cosmic ray masks generated by KcwiKit from three consecutive red channel exposures. The DAR-corrected cubes were fed into KCWISkywizard, which performed bad wavelength cropping, flux calibration, telluric corrections and further sky subtraction improvement with the Zurich Atmospheric Purge package \citep{2016MNRAS.458.3210S}. Exposures in each channel were cross-correlated, aligned with white-light images and drizzled onto a common spatial grid with \as{0.146} pixel scale in both directions \citep[][version 6.0 was used for drizzling]{Jacob10}.

BL and RL images were created by summing along the wavelength axis (see Supplementary Figure 6) and these both show the extension but no secondary source within it.
Spectra were extracted using a \as{1.0} diameter aperture for the central AGN and a \as{0.73} diameter aperture for the extended region. The extension spectra show no obvious features to indicate a secondary source. We conclude that the extension is intrinsic to the AGN.

\subsection*{Multiwavelength data}
\begin{figure*}
    \centering
    \includegraphics[width=\textwidth]{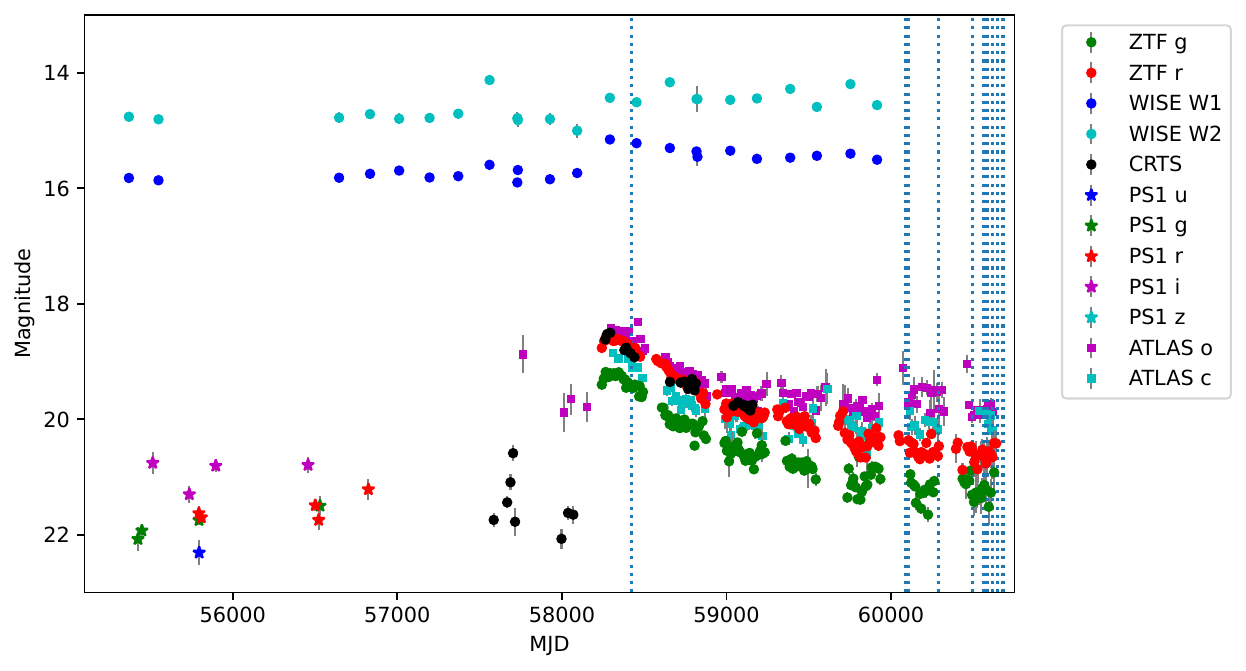}
    \caption{Photometric data for \agn\, from PS1, CRTS, ZTF, ATLAS, and WISE surveys, as indicated (see text for details). The dashed vertical lines indicate when spectra were obtained (see Supplementary Figure 3). Data are presented as observed values with one sigma measurement errors.}
    \label{fig:lightcurve}
\end{figure*}

\begin{figure*}
\centering
\includegraphics[width=0.8\textwidth]{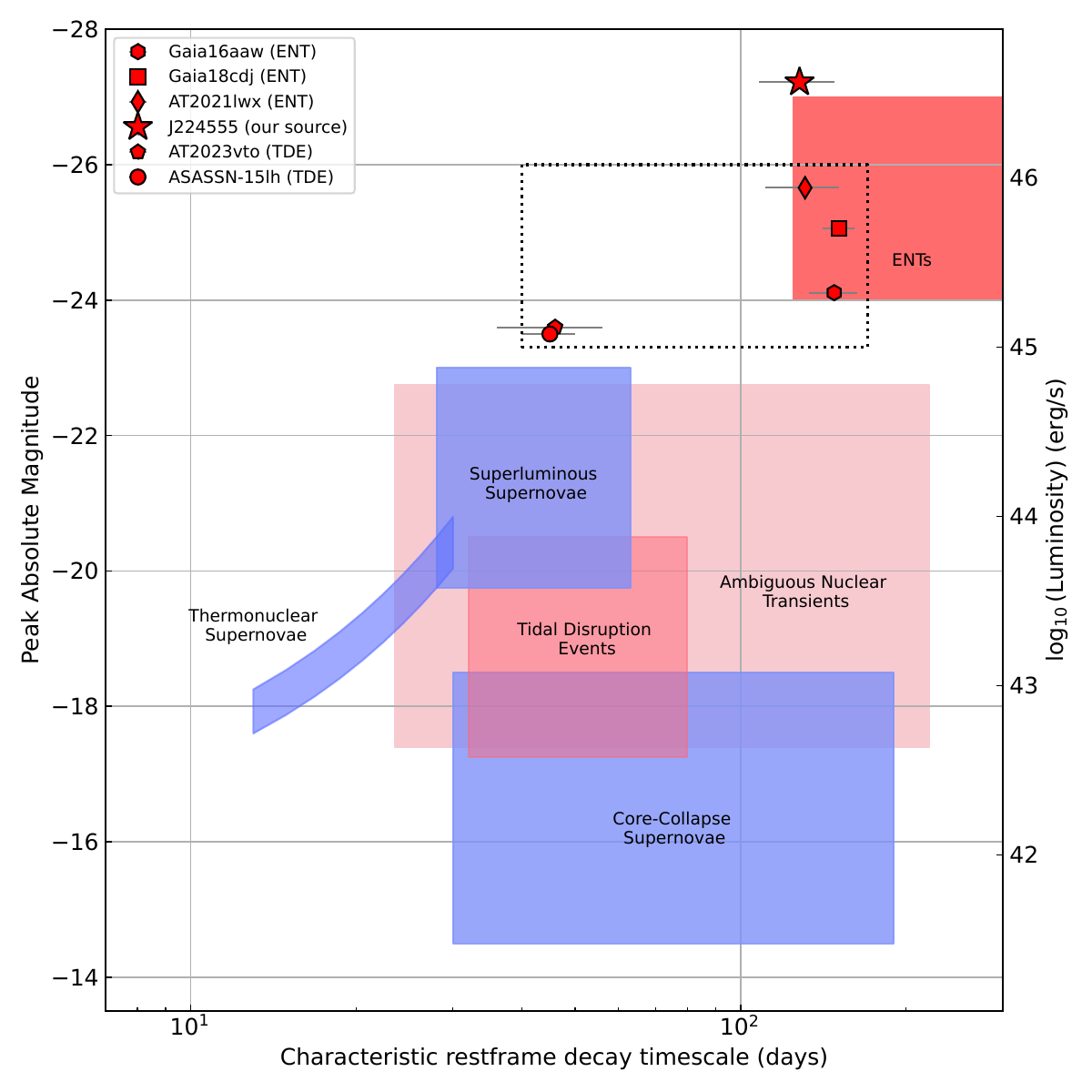}
    \caption{Peak absolute magnitude vs. restframe timescale to fade to 50\% peak flux. This shows the overlapping regions from \cite{Hinkle24} as well as the locations of the three ENTs published to date, and the brightest classical TDE to date (see text). The dotted boundary indicates a putative region containing intermediate mass TDEs. Data are presented as observed values with standard deviation uncertainties.}
    \label{fig:absmagtime}
\end{figure*}

\begin{figure*}
\centering
\includegraphics[width=\textwidth]{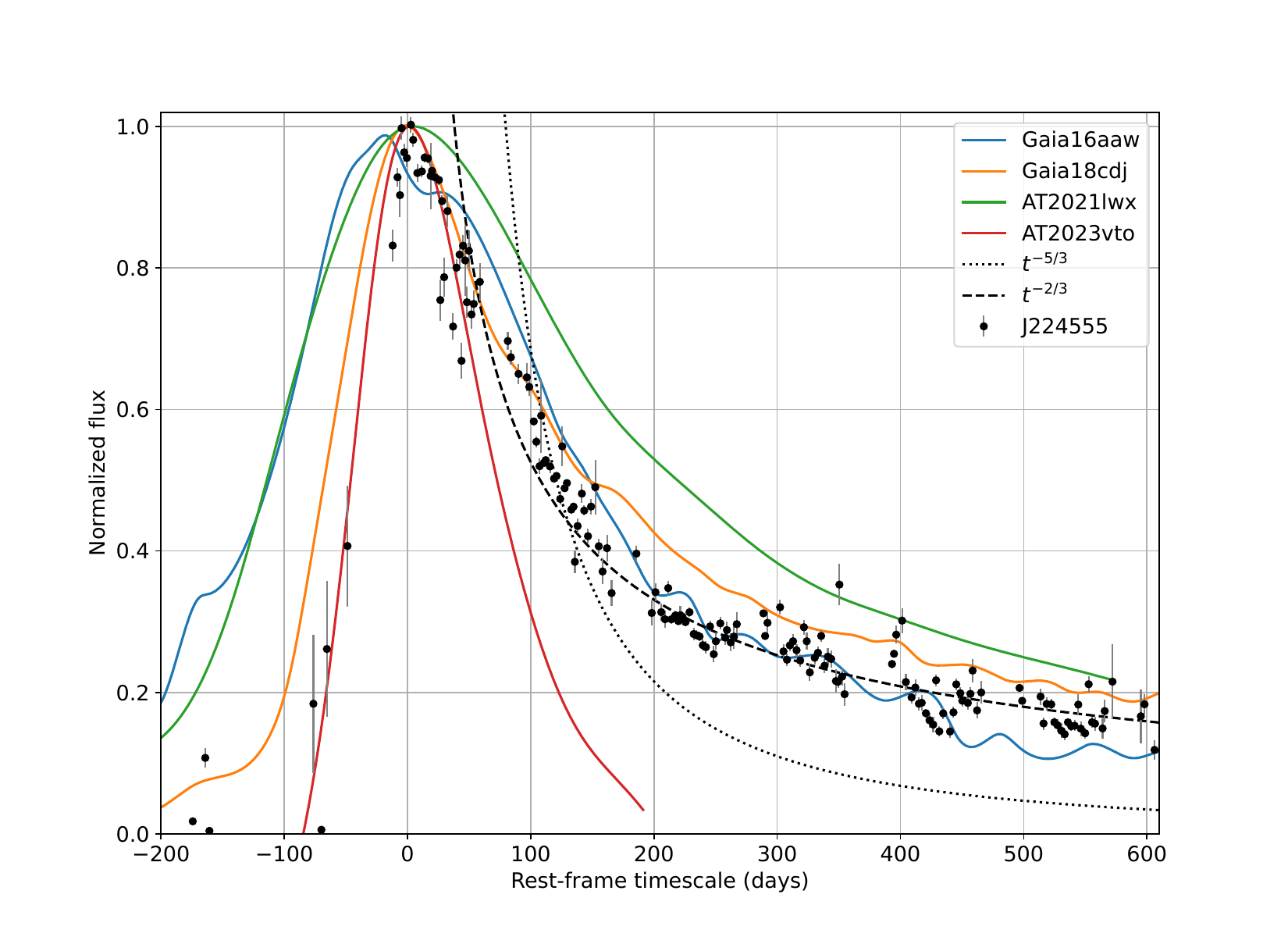}
    \caption{A comparison of the normalized light curve of \agn\ with ENTs, including AT2021lwx. The ENT light curves show Gaussian process fits to the photometric data with a Mat\'{e}rn-3/2 kernel. The dotted and dashed lines show the best-fit power law fits to the declining part of the light curve for -5/3 and -2/3 exponents, respectively. \agn\ shows a similar fast rise, like the superluminous TDE-He AT2023vto, but then a much slower decline, similar to ENTs. Data are presented as observed values with one sigma measurement errors.} 
    \label{fig:tde}
\end{figure*}

The Wide-field Infrared Survey Explorer \cite[WISE;][]{Wright10} and its extended mission, the Near-Earth Object WISE \cite[NEOWISE;][]{Mainzer14}, observed the entire sky in two mid-IR passbands, $W1$ (3.4 $\mu$m) and $W2$ (4.6 $\mu$m), from 2010 to 2011 (WISE) and 2014 to 2024 (NEOWISE) with a half-year cadence. The $W3$ (12 $\mu$m) and $W4$ (22 $\mu$m) passbands were also employed during the initial WISE phase. \agn\ is detected in the mid-IR by WISE with magnitudes $W1 = 15.913 \pm 0.048, W2 = 14.947 \pm 0.06, W3 = 11.818 \pm 0.253$ and $W4 = 9.101 \pm 0.499$. The corresponding WISE colors of $W1 - W2 = 0.966$ and $W2 - W3 = 3.129$ are consistent with an AGN \cite{Stern12}. The mid-IR light curve also shows an increase in flux at approximately the same epoch as the flare (see Fig.~\ref{fig:lightcurve}): this is consistent with related activity rather than some form of IR echo. The WISE $W1$ and $W2$ filters roughly correspond to slightly redward of $z$ and $J$ passbands in the restframe of \agn. Since the dusty torus typically dominates AGN emission at wavelengths longer than 1 $\mu$m, the WISE $W2$ band is still expected to probe dust emission.

Using a Gaussian process cross-correlation approach with a Mat\'{e}rn-3/2 kernel \cite{PozoNunez23}, we measure the observer frame time lags between ZTF $g$, ZTF $r$, WISE $W1$, and WISE $W2$ as: $g - r$ = $2.2 \pm 0.8$ days, $r - W1$ = $51.8 \pm 8.1$ days, $g - W1$ = $50.6^{+3.7}_{-2.1}$ days, and $r - W2$ = $53.4^{+8.2}_{-6.6}$ days.  Given the six-month cadence of the WISE survey, the uncertainties of the observed mid-IR lags likely have an additional, larger systematic contribution. The ZTF $r$ - WISE $W1$ restframe lag of 14 days is of the same order as the extended size of the TDE debris (11 days) in the models of \cite{Guillochon13, Guillochon14}. A reverberation
mapping-based model \cite[][]{Mandal24} gives an expected optical-IR lag of 500 -- 600 days for a bolometric luminosity between $10^{45.8}$ and $10^{46.6}$ erg s$^{-1}$ ($>$650 days for $L_{\rm bol} > 10^{47}$ erg s$^{-1}$), which is substantially longer than the measured lag. We infer that the transient IR emission is not reverberating flux from the inner edge of the dusty torus, but instead implies (roughly) co-spatial broadband emission related to the TDE.

The dust covering fraction in nuclear transients can be estimated as the ratio of the peak IR luminosity to the peak UV/optical luminosity \cite{Jiang21}.  The host emission is estimated from the pre-flare level and subtracted from the WISE light curves, and we correct for Galactic foreground extinction using the dust map of \cite{Schlegel98}. The flare peaks at $1.29 \times 10^{46}$ erg s$^{-1}$ in $W2$ which gives a dust covering fraction of 0.34 relative to the peak in the $r$-band light curve. This is consistent with AGN activity and other extreme nuclear transients \citep{Hinkle24}.

We obtained X-ray observations of \agn\, on UT 2024 August 14 with the X-Ray Telescope \cite[XRT;][]{Burrows05} onboard the Neil Gehrels Swift Observatory \cite{Gehrels04}. We analyzed the data  with HEASoft (v6.34) and the corresponding calibration files. Following standard data reduction procedures as outlined in \cite{Margutti13}, we do not find evidence for X-ray emission at the location of the transient in the 4.8\,ks exposure.  We infer a $3\,\sigma$ upper limit on the count-rate of $4.5\times 10^{-3}\,\rm{c\,s^{-1}}$ (0.3-10 keV). The neutral hydrogen column density along the line of sight is $1.0\times 10^{21}\,\rm{cm^{-2}}$ \citep{HI4PI}. For an assumed non-thermal spectrum $F_{\nu}\propto \nu^{-1}$, the corresponding 0.3-10 keV flux is $< 1.6\times 10^{-13}\,\rm{erg\, s^{-1}\, cm^{-2}}$ assuming no intrinsic absorption, corresponding to $L_X < 8.7 \times 10^{45}$ erg\, s$^{-1}$.


The source is undetected at radio wavelengths in the FIRST ($S_{\rm 1.4~GHz} <1.4$ mJy), NVSS ($S_{\rm 1.4~GHz} <2.5$ mJy), and VLASS ($S_{\rm 3~GHz}< 0.5$ mJy) radio surveys. It is detected with a total flux of $2.2 \pm 0.4$ mJy ($S_{\rm 120 - 168~MHz}$) in the LOFAR Two-metre Sky Survey DR2 \cite{Shimwell22}.
The NVSS and FIRST observations were taken pre-flare but the LOFAR and VLASS
are post-flare, LOFAR in 2019 and VLASS between 2019 and 2024. The post-flare optical/radio ratio (assuming a spectral index $\alpha \sim 0.7$) is more comparable with radio-loud than radio-quiet sources but the pre-flare constraints are too weak to tell whether the source was radio-loud at that time. In comparison to the jet associated with the jetted TDE Swift~J1644 \cite{Andreoni22}, the inferred post-flare spectral index for \agn\ is too steep and should be peaking in GHz frequencies. This strongly suggests that the radio emission is not related to the transient but to pre-flare activity. 
There is also no detected neutrino emission coincident with \agn\ from IceCube data.

Although we cannot measure the full bolometric output of \agn, we can place a lower limit to the total energy emitted by considering the individual passband host-subtracted light curves (see Supplementary Figure 1) and integrating from 100 days pre-peak to 650 days post-peak in the restframe. Using the CRTS + ZTF $r$-band directly gives a radiated energy of $8.4 \times 10^{53}$ erg and integrating a model fit to the data gives $8.9 \times 10^{53}$ erg. The WISE $W1$ and $W2$ host-subtracted light curves similarly give energies of $2.9 \times 10^{53}$ and $1.5 \times 10^{53}$ erg, respectively. Alternatively, integrating the parametric SED model fit (covering a restframe range of 0.89 -- 30.2 $\mu$m) to the spectra over the timeframe of the transient gives a total energy of $8.0 \times 10^{53}$ erg. \cite{Duras20} estimate an optical bolometric correction of $K_O \sim 5$ based on the $B$-band luminosity at 4400 \AA, which, from the NIRES spectrum, gives a bolometric luminosity at $t = 640$ days of $2.4 \times 10^{46}$ erg s$^{-1}$. Assuming this decays at the same rate as the optical emission, the flare produces a total electromagnetic energy of at least $10^{54}$ erg. It could be argued that since the host is an AGN, the host contribution is not constant over the duration of the flare. If we assume that it increases smoothly from the pre-flare level to the post-flare level, such that the level at $t = 650$ days in the restframe is the new host flux level, the total flux of the flare is only reduced by $\sim 10^{53}$ erg.

We can also estimate the accretion rate for the AGN from its Eddington ratio $\eta_{\mathrm{Edd}} = L_{\mathrm{bol}} / L_{\mathrm{Edd}}$, where $L_{\mathrm{Edd}}$ is the Eddington limit. Assuming a mass of $10^{8.5} M_\odot$ for the SMBH and $L_{\mathrm{bol}} = K_0 L_{4400}$, we estimate the preflare bolometric luminosity as $L_{\mathrm{bol}} = 2 \times 10^{45}$ erg s$^{-1}$ and $\log_{10} \eta_{\mathrm{Edd}} = -1.3^{+0.3}_{-0.4} $. This is at the lower end of the range for AGN in the redshift range $2.0 < z < 3.0$, $\log_{10}\eta_{\mathrm{Edd}} = -0.75 \pm 0.35 $ \cite{Kozlowski17}. The peak flare luminosity corresponds to $\eta_{\mathrm{Edd}} \sim 1$ and the current observed value has declined to $\eta_{\mathrm{Edd}} \sim 0.2$.

\subsection*{Identification as an AGN}
As mentioned above, the PS1 optical and WISE infrared colors of the pre-flare source, the position of the source in the BPT diagram, and potential pre-flare radio activity are all consistent with \agn\ being an AGN. The presence of forbidden oxygen lines with the observed flux ratios supports pre-flare AGN activity as they are only expected to change on multi-decadal timescales. Normal AGN variability is associated with a correlation between decreasing broad line width and increasing line luminosity, known as AGN breathing. Reverberation mapping and the virial theorem give this relationship as $\Delta v = \alpha L_{\rm line}^{-0.25}$ \cite{Cackett06}. \cite{Wang20} find that this effect is only strongly seen in Balmer emission lines ($\alpha \sim -0.25$); \mgii\ shows effectively no correlation ($\alpha \sim 0$),  and an anti-correlation $(\alpha > 0)$ is seen in  \civ, \ciii, and Si~IV.  In flaring AGN, the continuum luminosity can increase by orders of magnitude on short timescales without an immediate response from the broad line region (BLR). In fact, line widths may even increase with increasing luminosity due to the appearance of new kinematic high-velocity non-virial gas components, such as winds or shocks, or line production from ionized gas outside the BLR. Both changing-look AGN \cite[e.g.][]{MacLeod19, Sheng17,MacLeod16} and TDEs \cite[e.g.][]{Graham17, Ricci20} are seen to show broader lines when brighter. Supplementary Figure 7 shows the relation between broad line width and line luminosity for \Lya\, \civ\, \ciii\, and \mgii\ for \agn. The latter three lines show no correlation or an anti-correlation and \Lya\ may show a weak correlation. This behavior is consistent with normal AGN activity.

\subsection*{Light curve properties}
The rise and decline times of transients are characterized in several ways in the astronomical literature, depending on subdomain practices: empirically according to some prescription, e.g., e-folding,  half peak, or from a particular model fit. 
\agn\ shows an e-folding rise time of 49 days and an e-folding decay time of 167 days in the restframe. Half-peak times are $t_{\rm 1/2, rise} = 30$ days and $t_{\rm 1/2, decay} = 112$ days in the restframe. A Gaussian-rise exponential decay model, typically used to model TDEs, gives characteristic timescales of 31 days for the Gaussian rise and 132 days for the decay in the restframe.

An alternate approach from the characterization of X-ray solar and optical stellar flares \cite{Mendoza22, Gryciuk17} is to relate the shape of the observed light curve to the time profile of energy release. The release and deposition of energy during a flare is described by a function, $g(x)$, and the dissipation of released energy by a monotonically decreasing function, $h(x)$. The available energy at a moment, $t$, following an interval, ${\mathrm d}x$, is then ${\mathrm d}f(t) = g(x) h(t - x) {\mathrm d}x$ and the overall event time profile is given by: 
\[ f(t) = \int_{-\infty}^t g(x) h(t-x) {\mathrm d}x \]

\noindent 
A Gaussian term can be used to account for the impulsive heating during the rise phase of the flare with characteristic timescale, $t_g$, $g(x) = A e^{-(x - t_0)^2 / t_g^2}$, and an exponential for the decay phase. A more flexible parameterization, however, is to use
a double exponential to account for possible rapid and gradual cooling phases, adopting respective timescales $t_{rc}$ and $t_{sc}$ which may reflect thermal and non-thermal dissipation modes, respectively: $h(x) = F_1 e^{-x/t_{rc}} + F_2 e^{-x/t_{sc}}$. $F_1$ and $F_2 = 1 - F_1$ account for the relative contribution of each exponential term.
For AGN, we also want to allow for the possibility that the flare may affect the background level of activity in the AGN so that the post-flare luminosity of the AGN would be less or lead to a higher accretion rate and an increased post-flare luminosity.  The full flare profile is then:
\begin{eqnarray*}
f(t) & = & f_0 + \frac{\sqrt{\pi}At_g}{2} (F_1h(t, t_0, t_g, t_{rc}) + F_2h(t,t_0,t_g, t_{sc})), \,\, t \le t_0 \\
  & = & f_1 + \frac{\sqrt{\pi}A't_g}{2} (F_1h(t, t_0, t_g, t_{rc}) + F_2h(t,t_0,t_g, t_{sc})), \,\, t > t_0 \\
h(t, t_0, t_g, t_c) & = & \exp \left[ \frac{t_0 - t}{t_c} + \frac{t_c^2}{4 t_g^2}  \right] erfc \left(\frac{t_0 - t}{t_g} + \frac{t_g}{2 t_c} \right)
\end{eqnarray*}

\noindent 
where $erfc(t)$ is the complementary error function.

A fit to the restframe flux light curve of \agn\ (see Supplementary Figure 4) gives a characteristic rise time of $t_g = 39$ days and decay times of $t_{rc} = 47$ days and $t_{sc} = 141$ days, respectively, with $F_1 = 0.29$. We note that there is some substructure in the fading part of the light curve but there are no significant indications of repeat flaring, as might be expected from a partial TDE or periodicity. The variability appears consistent with regular AGN behavior.



\backmatter
\section*{Declarations}
The authors declare no competing interests.

\bmhead{Data Availability}
All photometry and spectroscopy used for \agn\ is available in a public Github repository (\url{www.github.com/doccosmos/superman}). All spectra and photometry for other sources in figures have been taken from public archives.

\bmhead{Code Availability}
KCWISkyWizard is an open source tool available at: \url{github.com/zhuyunz/KSkyWizard}



\bmhead{Acknowledgements}

We thank the two referees for their helpful comments.

MJG acknowledges support by the US National Science Foundation (NSF) through grant AST-2108402. BM \& KESF are supported by NSF AST-1831415, NSF AST-2206096 and Simons Foundation Grant 533845. The work of DS was carried out at the Jet Propulsion Laboratory, California Institute of Technology, under a contract
with the National Aeronautics and Space Administration (80NM0018D0004). SGD acknowledges generous support from the Ajax Foundatio. PMV acknowledges the support from the DFG via the Collaborative Research Center SFB1491 \textit{Cosmic Interacting Matters - From Source to Signal}. PW acknowledges support from the Science and Technology Facilities Council (STFC) grant ST/Z510269/1. 

Based on observations obtained with the Samuel Oschin Telescope 48-inch and the 60-inch Telescope at the Palomar Observatory as part of the Zwicky Transient Facility project. ZTF is supported by the National Science Foundation under Grants No. AST-1440341 and AST-2034437 and a collaboration including current partners Caltech, IPAC, the Oskar Klein Center at Stockholm University, the University of Maryland, University of California, Berkeley, the University of Wisconsin at Milwaukee, University of Warwick, Ruhr University Bochum, Cornell University, Northwestern University and Drexel University. Operations are conducted by COO, IPAC, and UW.

This publication makes use of data products from the Wide-field Infrared Survey Explorer, which is a joint project of the University of California, Los Angeles, and the Jet Propulsion Laboratory/California Institute of Technology, and NEOWISE, which is a project of the Jet Propulsion Laboratory/California Institute of Technology. WISE and NEOWISE are funded by the National Aeronautics and Space Administration.

\bmhead{Author Contributions}
M.J.G. reduced the optical data, conducted the spectral and photometric analysis, and is the primary author of this manuscript. B.M. and K.E.S.F. contributed to the theoretical interpretations of the source. D.S. obtained some of the optical spectra and contributed to the text. M.C. provided text for the supernova section. A.J.D. initially identified the source in CRTS data and provided the forced CRTS light curve and text. Y.D. and M.Koss provided the KCWI data and text description. M. Kasliwal and S.R. provided the NIRES data and text description. R.M. provided the SWIFT data and text description. J.S. provided the radio data and text description. S.G.D. is the co-PI of NSF AST-2108402. P.W. and P.M.V. contributed to reviewing the text. E.C.B. is the ZTF Survey Scientist. T.X.C. and S.L.G. provided IPAC data support for ZTF. S.R.K. was the PI of ZTF. A.M. is the machine learning lead for ZTF.

\bmhead{Competing Interests}

\clearpage
\begin{table}
\caption{Spectroscopic observations of \agn. The final column adopts a peak date of MJD = 58295 (i.e., UT 2018 June 26).}
\label{tab:spectra}
\begin{tabular}{lcccc}
\toprule
Date (UT) & Telescope & Instrument & Exposure time (s) & Restframe days post-peak \\
\midrule
2018 November 2$^*$ & P200 & DBSP & $2 \times 900$ & 36 \\
2023 May 21 & P200 & DBSP & $2 \times 1200$ & 505 \\
2023 July 13 &  Keck-I & LRIS & $2 \times 1200$ & 509 \\
2023 December 9$^*$ & Keck-I & LRIS & $2 \times 1200$ & 562 \\
2024 July 4 & Keck-I & LRIS & 900 & 621 \\
2024 September 11 & P200 & DBSP & $3 \times 1200$ & 640 \\
2024 September 11 & Keck-II & NIRES & $6 \times 300$ & 640 \\
2024 October 1 & Keck-I & LRIS & 1200 & 649 \\
2024 November 1 & Keck-I & LRIS & $2 \times 300$ & 659 \\
2024 December 4 & Keck-II & KCWI & 2400/1800 & 670 \\
2025 January 1 & Keck-I & LRIS & $2 \times 1200$ & 678 \\
\bottomrule
\end{tabular}
* Note: Poor observing conditions.
\end{table}

\begin{table}
\caption{Transient sources used for comparison with \agn: slow declining superluminous supernovae (SLSNe), candidate pair-instability supernovae (PISNe), and extreme nuclear transients (ENTs)/intermediate mass TDEs.}
\label{tab:comp}
\begin{tabular}{lcccccc}
\toprule
ID & Other ID & RA & Dec & Redshift & Classification & Ref. \\
\midrule 
ZTF18aavskep & AT2018bwr & 15:28:26.17 & +08:48:22.20 & 0.046 & SLSN-II (IIn) & (1) \\
ZTF18acbvhfl & AT2018hse & 03:50:06.28 & -11:00:12.65 & 0.130 & SLSN-II & (1) \\
ZTF19aafnend & AT2019bhg & 16:59:26.21 & +46:47:56.94 & 0.1203 & SLSN-II (IIn) & (1) \\
ZTF19abcejsg & AT2019jyu & 00:22:38.69 & -00:09:28.74 & 0.134 & SLSN-II (IIp) & (1) \\
ZTF19abpvbzf & AT2019npx & 20:32:52.07 & -12:51:00.32 & 0.057 & SLSN-II & (1) \\
ZTF20acusylb & AT2020abku & 14:00:51.22 & -00:11:07.98 & 0.0818 & SLSN-II (IIn) & (1) \\
\midrule
SN 2007bi & - & 13:19:20.19 & 08:55:44.29 & 0.1279 & PISN & (2) \\
ZTF18acenqto & AT2018ibb & 04:38:56.95 &  -20:39:44.10 & 0.166 & PISN & (2) \\ 
\midrule
Gaia16aaw & AT2016dbs & 04:11:57.03 & -42:05:30.84 & 1.03 & ENT & (4)\\
Gaia18cdj & AT2018fbb & 02:09:48.14 & -42:04:37.02 & 0.93747 & ENT & (4) \\
ZTF20abrbaie & AT2021lwx & 21:13:48.41 & +27:25:50.4 & 0.995 & ENT & (4) \\
ZTF23abcvbqq & AT2023vto & 00:24:34.71 & 47:13:21.55 & 0.4846 & TDE-He & (5) \\
\bottomrule
\end{tabular}
References: (1) Pessi et al. 2024; (2) Gal-Yam et al. 2009; (3) Shulze et al. 2024; (4) Hinkle et al. 2024; (5) Kumar et al. 2024
\end{table}

\clearpage

\clearpage

\clearpage

\clearpage
\section*{Supplementary Text}\label{sec:MainText}

\section*{Discussion}

We consider here \agn\ in terms of the most plausible remaining explanations.

\subsection*{The case for a nuclear supernova}
\label{sec:sne}
As discussed above, we disfavour an individual supernova on the basis of energetics. \agn\ is $\sim 2(3)$ orders of magnitude more luminous than typical super-luminous supernovae (core-collapse supernovae) or SLSN(CCSN) (see also Supplementary Figures~\ref{fig:snspeccomp} and \ref{fig:sne} for comparison of \agn\ to supernova spectra and lightcurves). The maximum possible release of energy from a supernova is expected to be $\mathcal{O}(10^{53}{\rm erg})$ \citep{Woosley21,Renzo24}, or an order of magnitude less energy than \agn. A more detailed discussion of these considerations follows below.

The presence of a nuclear star cluster (NSC) around an SMBH suggests we should occasionally see the endpoints of stellar evolution as a source of transient phenomena in AGN. Most observed CCSNe result from the explosion of stars more massive than $\sim 8\, M_{\odot}$ and output total energy in the range $E_{\rm SN} \sim 10^{49-51}\, {\rm erg}$, depending on supernova type and inferred mass \citep{Chevalier11}. The new class of SLSNe \citep{GalYam19, Pessi24} are an order of magnitude more energetic than CCSNe and may be powered by the most massive stars ($M_{\ast} \gtrsim 40\, M_{\odot}$). If \agn\ is powered by a supernova, it is  orders of magnitude more luminous than typical CCSN and SLSN (Supplementary Figure~\ref{fig:sne}). 

Massive stars, the sources of CCSN, are rare in the field, but could be more common in NSCs (particularly those interacting with AGN) if the initial mass function (IMF) of stars is biased towards top-heavy as predicted \citep{Bartko10,Neumayer20}. In addition, stars that are fully embedded in AGN disks may evolve very differently from stars in the field and quickly grow supermassive \citep{Cantiello21, Jermyn22}. Such stars then collapse violently, depending on the surrounding gas density \citep{Fabj24}. Massive stars that explode in AGN disks will have their luminosity boosted as the kinetic energy of the explosion is turned into radiation by surrounding material \citep{Grishin21}.

Sufficiently massive stars ($M_{\ast} \sim 140-260M_{\odot}$) are expected to undergo pair-instability (or pulsation pair-instability) in their cores at the ends of their life, which initiates a pair instability supernova (PISN) \citep{Renzo24}.  While there is no direct observational evidence for a PISN yet, potential candidates include, e.g., SN 2007bi \cite{GalYam09} and SN 2018ibb \cite{Schulze24}, each of which radiated a total energy of $\mathcal{O}(10^{51-52}\, {\rm erg})$. Candidate PISN light curves are characterized by a slow rise and decline, lasting several months, longer than typical CCSN. PISN spectra are expected to show broad absorption features from highly ionized elements, e.g., oxygen and silicon, but an absence of hydrogen and helium. The most massive stars ($M_{\ast} \gtrsim 260M_{\odot}$) do not undergo PISN and are expected to core collapse directly to black holes without an explosion, liberating relatively small amounts of energy \citep{Woosley02,Woosley21,Renzo24}. Unless our present understanding of how very massive stars explode significantly underestimates the energy released, it is not possible for a supernova to generate the energy observed in \agn.

Several emission lines in the spectra of \agn\ have P~Cygni profiles (see Supplementary Figure~\ref{fig:pcyglines}) which could be indicative of supernovae where an expanding shell interacts with dense circumstellar material. The velocity widths (FWHM) of the P Cygni profiles in \agn\ and their offsets are only $\sim 2000$ km s$^{-1}$, which is at the low end of the typical range for a SLSNe. We note, however, that the absorption features show no or little sign of evolution over the $\sim$650 restframe days since peak, suggestive of absorption associated with a distant screen of material and not the explosive event itself. SLSNe spectra should also evolve into a nebular phase with an appropriate set of emission lines after 2 - 4 e-folding times \cite{Nicholl19} which is about now. No such changes have yet been observed in \agn.

\subsection*{The case for gravitational lensing}
The energetics of the flare are difficult to reconcile with most known classes of transient, although there is some ambiguity about the mechanism(s) driving the most luminous transients. To date, the most luminous reported supernova has a peak absolute magnitude of -23.5 \cite[$2 \times 10^{45}$\, erg\, s$^{-1}$; ASASSN-15lh/SN 2015L;][]{Dong16} and the most luminous reported classical TDE has a peak absolute magnitude of -23.6 (peak bolometric luminosity of $8 \times 10^{44}$\, erg\, s$^{-1}$ [40]). \agn\ would represent a boosting of the peak luminosity of these by a factor of at least 30 and a much larger factor for a more representative, i.e., average, transient of either class. The most extreme AGN flares are also seen to have typical total energies in the range $10^{51} - 10^{52}$ erg \cite{Graham17} compared to $\sim10^{54}$ erg for \agn. 

Gravitational lensing by a massive body along the line of sight to \agn\ offers a potential solution: the restframe energetics of the flare may still be extreme but fall within a statistically realizable range boosted by a gravitational magnification factor, $\mu$, to the observed values. The range of magnification necessary is also moderate compared to extreme magnification events ($\mu > 1000)$ that have been observed \cite[e.g.,][]{Diego19}.

KCWI imaging of \agn\ shows no indication of multiple components so the angular separation between multiple lensed images would be smaller than the \as{0.147} angular resolution of KCWI. The detection of lensing would then be on the basis of magnification. For unresolved sources, the total flux as a function of time consists of the addition of the fluxes of each lensed image at their arrival times, $t_i$, multiplied by their magnification, $\mu_i$: $F_T = \sum_i^{N_m} F(t_i)\mu_i $. 

\cite{SaguesCarracedo24} simulated the detectability of lensed supernova, both thermonuclear (Ia) and core collapse, in regular host galaxies to a redshift of $z = 1.5$. We have run their simulation code, LENSIT (\url{github.com/asaguescar/LENSIT}), assuming a detectability redshift limit of $z = 2.5$ for both SLSNe-II and TDE populations. We employ the `nugent-sn2n' template for SLSNe-II with an absolute magnitude distribution given by a Gaussian $M_B \sim {\cal N}(-20.5, 0.5)$ and a local rate of $R_{\rm loc} = 213$ Gpc$^{-3}$ yr$^{-1}$ \cite{Pessi24}. We generated 100,000 unresolved lensed SLSNe-II light curves with a singular isothermal ellipsoid model and external shear for the lens galaxy. The code uses the actual ZTF observing logs to cast these as they would have been observed by the survey, complete with cadence, error bars, and wavelength coverage. Identifiable sources are defined as having at least five detections around the SN peak and a peak absolute magnitude brighter than -19.5.

This exercise identified 25 simulated light curves comparable to ENTs in peak apparent absolute magnitude $(M < -24)$, including five as luminous or more luminous than \agn. However, none match in peak apparent brightness and only three of the simulated supernovae are at $z > 2$. The closest simulation is a quad lens system with the source at $z = 2.35$ and the lens at $z = 1.43$ which peaks at $m = 20.5\, (M = -25.5)$. This is close to the limit of detectability for ZTF. Supplementary Figure~\ref{fig:sne} also shows that the normalized light curve of \agn\ is broadly consistent with those of about 5\% of SLSNe-II. We also find single epoch spectral similarities with $\sim 10$\% of SLSNe-II spectra, and the observed color evolution is comparable post-peak to observed SLSNe-II. The unlensed source would therefore be representative of about $\sim 1$\% of the SLSNe-II population. This means that the chance 
of \agn\ being a gravitationally lensed SLSNe-II at $z \sim 2.5$ is less than $10^{-5}$. 

A similar analysis shows that a lensed TDE, with a background rate less than SLSNe, is even less likely. We have also already argued (see above) that the odds of self-lensing are vanishingly small.

\subsection*{Other plausible explanations}
\cite{Lipunova24} propose ionization instabilities in the accretion disk as a mechanism to produce giant flares in AGN. A transition from a cool, neutral disk state to a hot, ionized state leads to super-Eddington accretion and strong outflows. The resultant flares are similar to TDEs but with potentially greater power and longer duration. They are most likely to occur in SMBHs with a truncated disk, where the inner disk is replaced by a radiatively inefficient accretion flow. The flare's magnitude and duration depend on the initial truncation radius and the subsequent advection-dominated state of the disk, with the turbulence parameter $\alpha$ growing in the ionized disk. Such events should be more prevalent around more massive SMBHs, particularly those with $M > 10^9 M_\odot$, but they may also be associated with the tidal disruption of a sufficiently massive star in a lower mass SMBH.

We have assumed that the flare emission is isotropic, particularly in estimating the energy associated with it. It may be possible instead that we are observing beamed or collimated emission associated with some transient event in an AGN disk. The integrated energy released ($10^{54}$ erg from just the optical light curve) comes from an initial event and 650 days of decay rather than 90\% released in the first week, say. The energy source is still ongoing after 650 days and this would need the whole flare to be strongly beamed over the full duration but the lack of any detectable X-ray or associated radio response after 600 days argues that there is no current strongly beamed component. The measurable time lag to the IR response also tells us that there is at least some initial isotropic component. Jetted TDEs \cite[e.g.,][]{Andreoni22} also have featureless optical spectra, very different from the AGN spectrum seen here and arguing against \agn\ being a jetted source.

\renewcommand{\tablename}{Supplementary Table}
\clearpage
\begin{table}
\caption{Transient sources used for comparison with \agn: slow declining superluminous supernovae (SLSNe), candidate pair-instability supernovae (PISNe), and extreme nuclear transients (ENTs)/intermediate mass TDEs.}
\label{tab:comp}
\begin{tabular}{lcccccc}
\toprule
ID & Other ID & RA & Dec & Redshift & Classification & Ref. \\
\midrule 
ZTF18aavskep & AT2018bwr & 15:28:26.17 & +08:48:22.20 & 0.046 & SLSN-II (IIn) & \cite{Pessi24} \\
ZTF18acbvhfl & AT2018hse & 03:50:06.28 & -11:00:12.65 & 0.130 & SLSN-II & \cite{Pessi24} \\
ZTF19aafnend & AT2019bhg & 16:59:26.21 & +46:47:56.94 & 0.1203 & SLSN-II (IIn) & \cite{Pessi24} \\
ZTF19abcejsg & AT2019jyu & 00:22:38.69 & -00:09:28.74 & 0.134 & SLSN-II (IIp) & \cite{Pessi24} \\
ZTF19abpvbzf & AT2019npx & 20:32:52.07 & -12:51:00.32 & 0.057 & SLSN-II & \cite{Pessi24} \\
ZTF20acusylb & AT2020abku & 14:00:51.22 & -00:11:07.98 & 0.0818 & SLSN-II (IIn) & \cite{Pessi24} \\
\midrule
SN 2007bi & - & 13:19:20.19 & 08:55:44.29 & 0.1279 & PISN & \cite{GalYam09} \\
ZTF18acenqto & AT2018ibb & 04:38:56.95 &  -20:39:44.10 & 0.166 & PISN & \cite{Schulze24} \\ 
\midrule
Gaia16aaw & AT2016dbs & 04:11:57.03 & -42:05:30.84 & 1.03 & ENT & [13] \\
Gaia18cdj & AT2018fbb & 02:09:48.14 & -42:04:37.02 & 0.93747 & ENT & [13] \\
ZTF20abrbaie & AT2021lwx & 21:13:48.41 & +27:25:50.4 & 0.995 & ENT & [13] \\
ZTF23abcvbqq & AT2023vto & 00:24:34.71 & 47:13:21.55 & 0.4846 & TDE-He & \cite{Kumar24} \\
\bottomrule
\end{tabular}
\end{table}

\renewcommand{\figurename}{Supplementary Figure}
\clearpage
\begin{figure*}
    \centering
    \includegraphics[height = 2.4in]{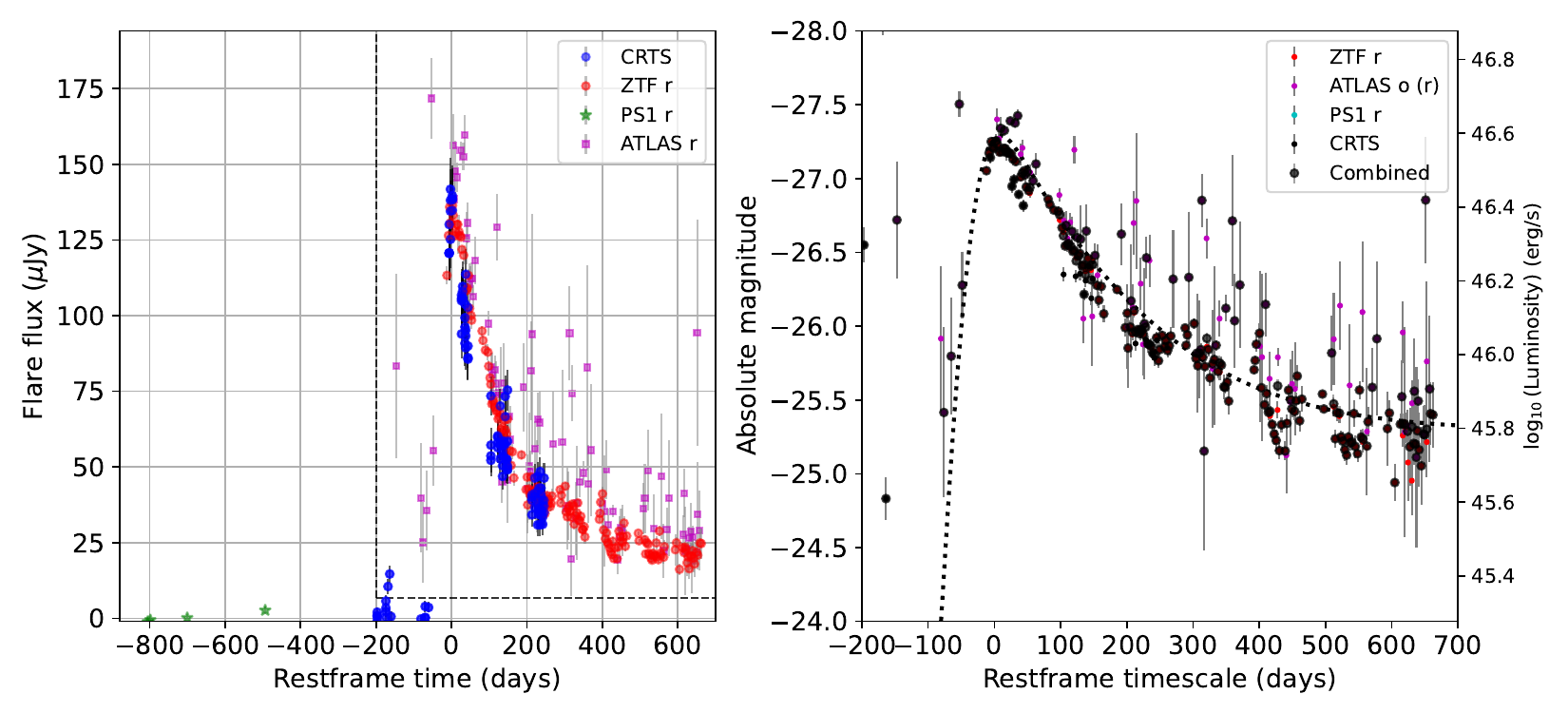}
    \caption{Forced photometric data in flux space for the flare in \agn\, from PS1, CRTS, ZTF, and ATLAS. CRTS and ATLAS $o$-band data have been transformed to PS1 $r$-band for direct comparison. Host contributions have been removed from the CRTS and PS1 data. The right panel highlights the region defined by the dashed lines in magnitude space. The dotted line indicates a best fit model to the combined data for the flare. The model \citep{Mendoza22} describes the time profile of energy release in the flare and features a Gaussian rise to account for impulsive heating and a double exponential to account for rapid and gradual cooling phases. Data are presented as observed values +/- one sigma measurement errors.}
    \label{fig:flare}
\end{figure*}

\clearpage
\begin{figure*}[h]
\centering
\includegraphics[width=\textwidth]{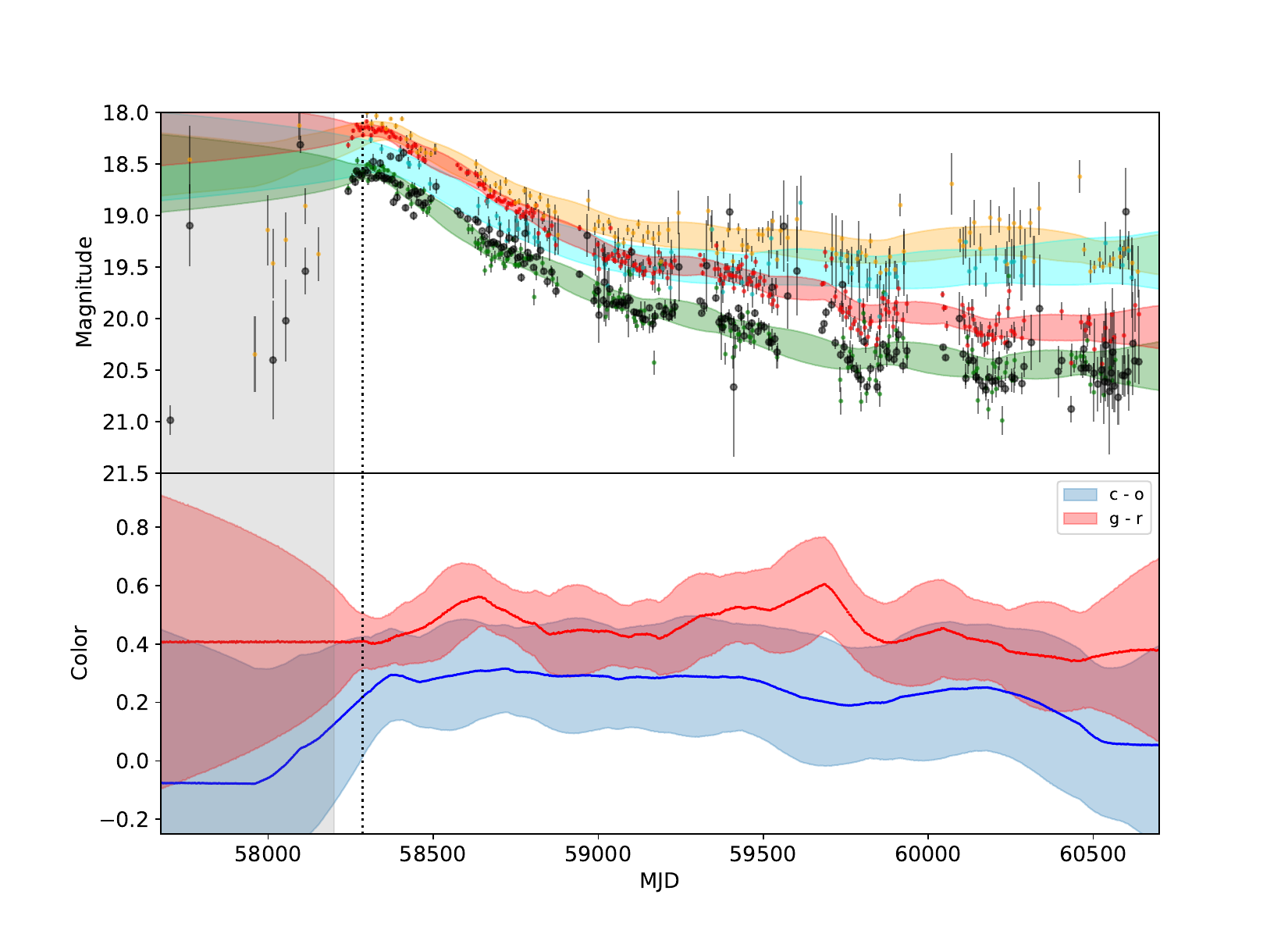}
    \caption{Multiband photometric data for \agn. (Top) The apparent magnitude evolution of \agn\ from ZTF $g$- (green) and $r$-bands (red) and ATLAS $c$- (cyan) and $o$-bands (orange). The black points show the combined data set calibrated to the $r$-band. The color bands indicate Gaussian process regression fits to the individual filter light curves and their variance. The vertical dotted line indicates the peak of the flare at MJD = 58295 days and the grayed region indicates the pre-peak time period when there is insufficient data for the Gaussian process and the fit is extrapolated. (Bottom) The color evolution of \agn. Minimal color evolution (i.e., statistically consistent with zero evolution) from the flare during its cooling phase. Data are presented as observed values +/- one sigma measurement errors.}
    \label{fig:color}
\end{figure*}

\clearpage
\begin{figure*}
    \includegraphics[width=\textwidth]{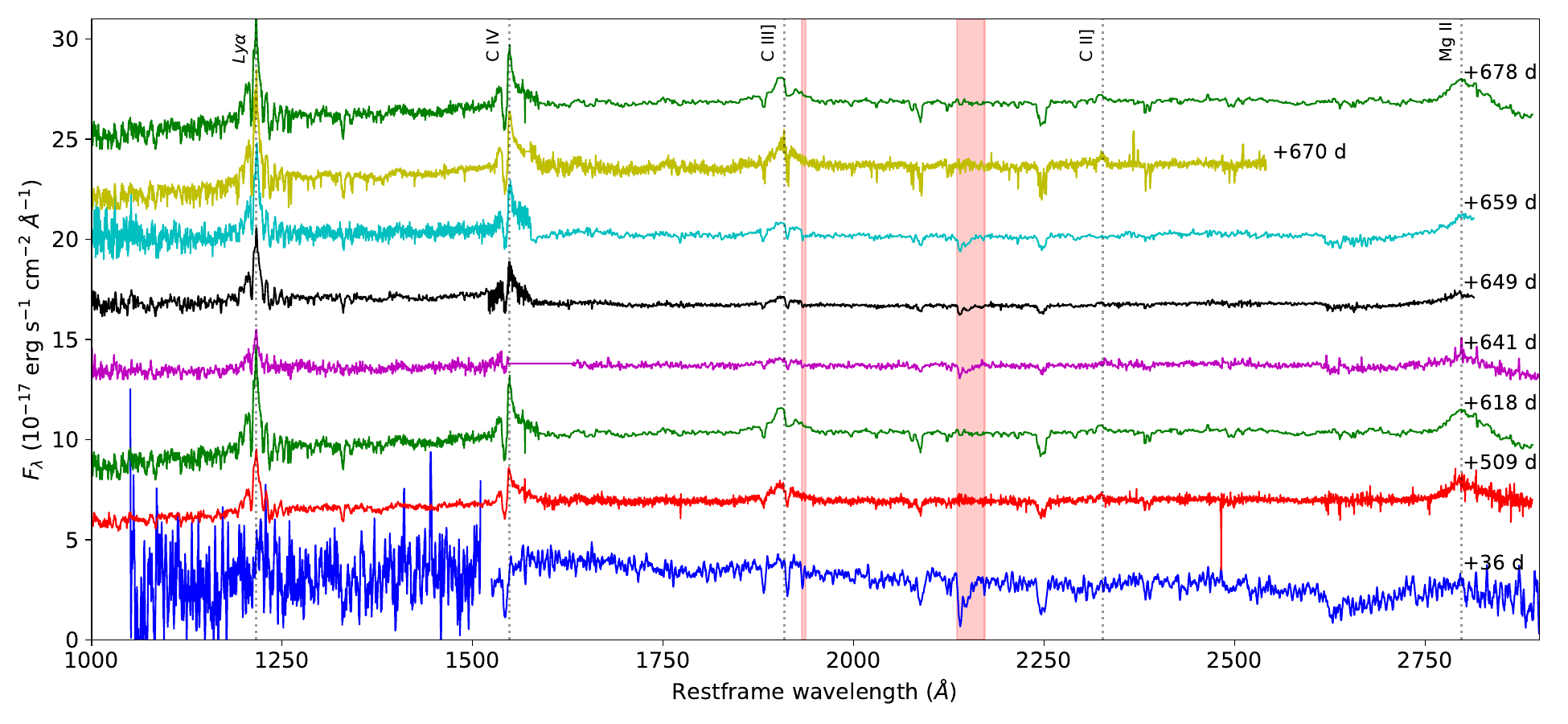}
    \includegraphics[width=\textwidth]{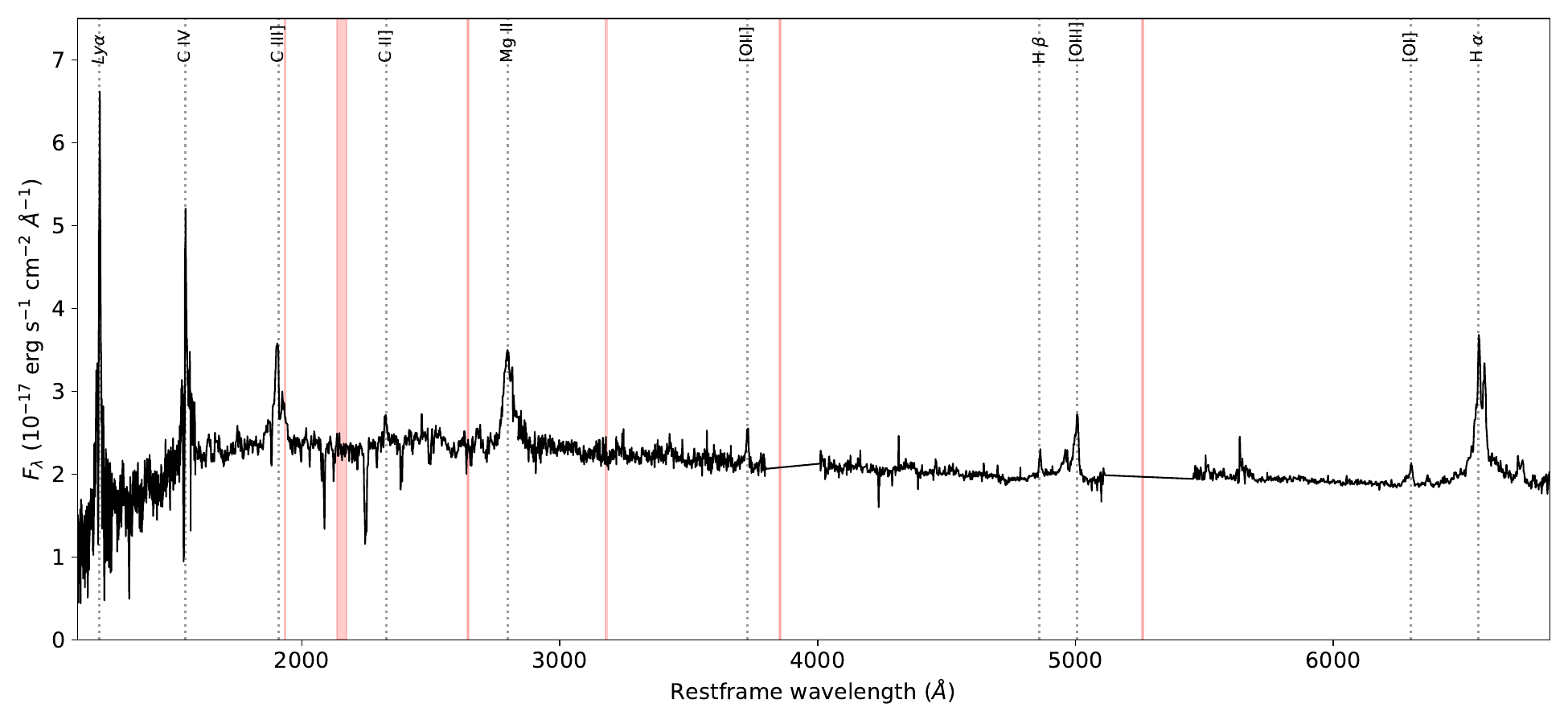}
    \includegraphics[width=\textwidth]{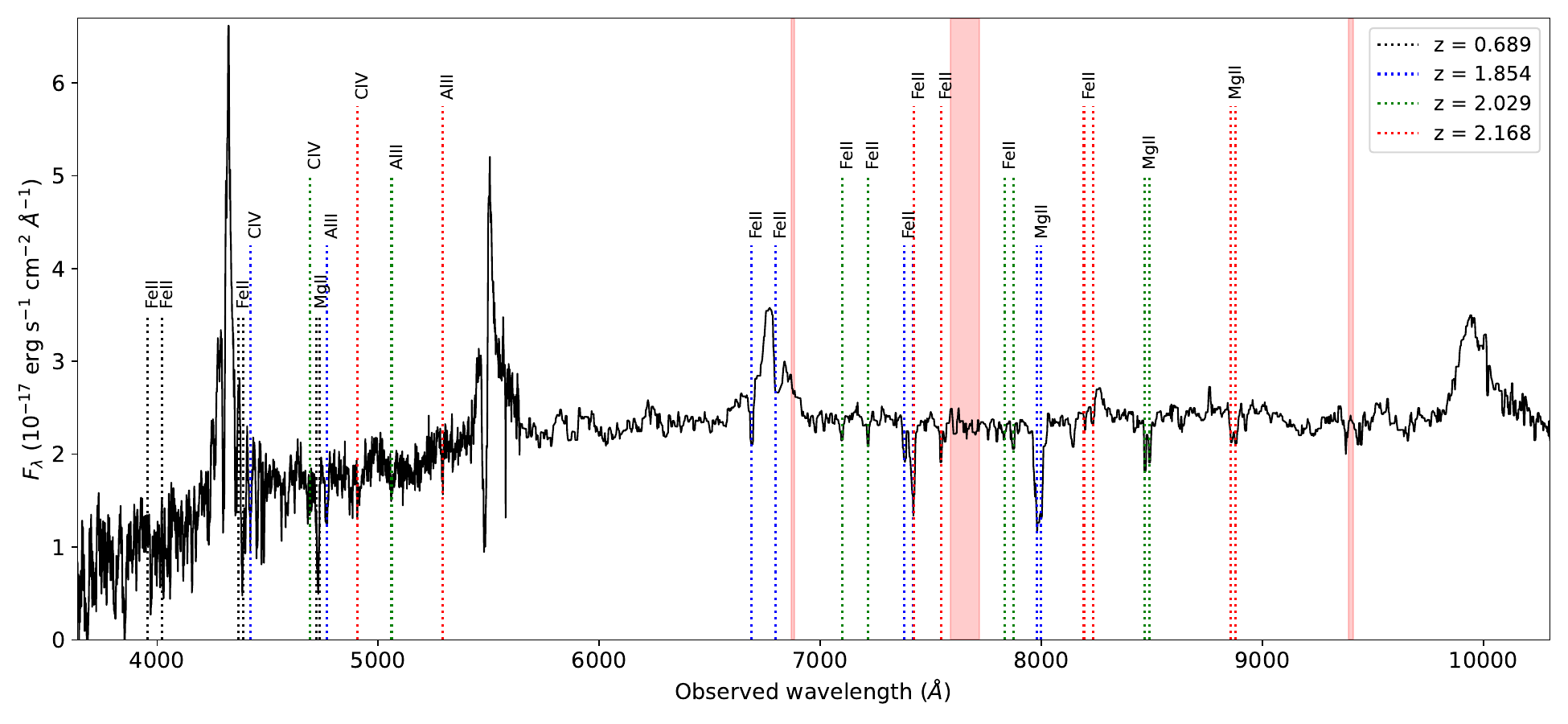}
    \caption{Spectroscopic data for \agn. (Top) Multiepoch optical spectra of \agn. The spectra are offset relative to each other for clarity. (Middle) A composite spectrum of \agn\ combining LRIS and NIRES data. (Bottom) Optical spectrum highlighting absorption systems along the line-of-sight at $z =$ 0.689, 1.854, 2.029 and 2.168. The shaded red regions indicate the locations of the telluric A and B bands.}
    \label{fig:spectra}
\end{figure*}

\clearpage
\begin{figure*}
 \centering
 \includegraphics[width=\textwidth]{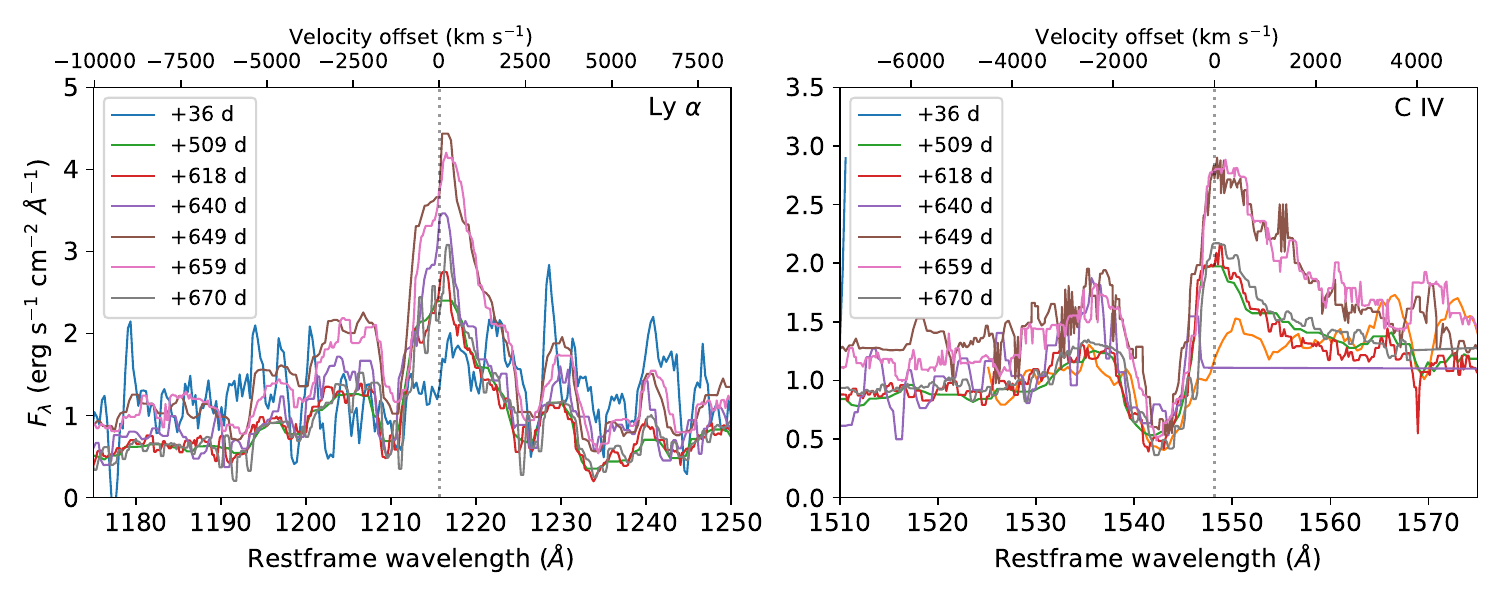}
    \caption{Zoom in of the strongest broad UV emission lines, \Lya and \civ, at different epochs. Clear P-Cygni-like absorption features can be seen in the spectra. The spectra have been scaled to have the same median flux over the restframe wavelength range 2150 -- 2250 \AA.}
    \label{fig:pcyglines}
\end{figure*}

\clearpage
\begin{figure*}
 \centering
 \includegraphics[width=\textwidth]{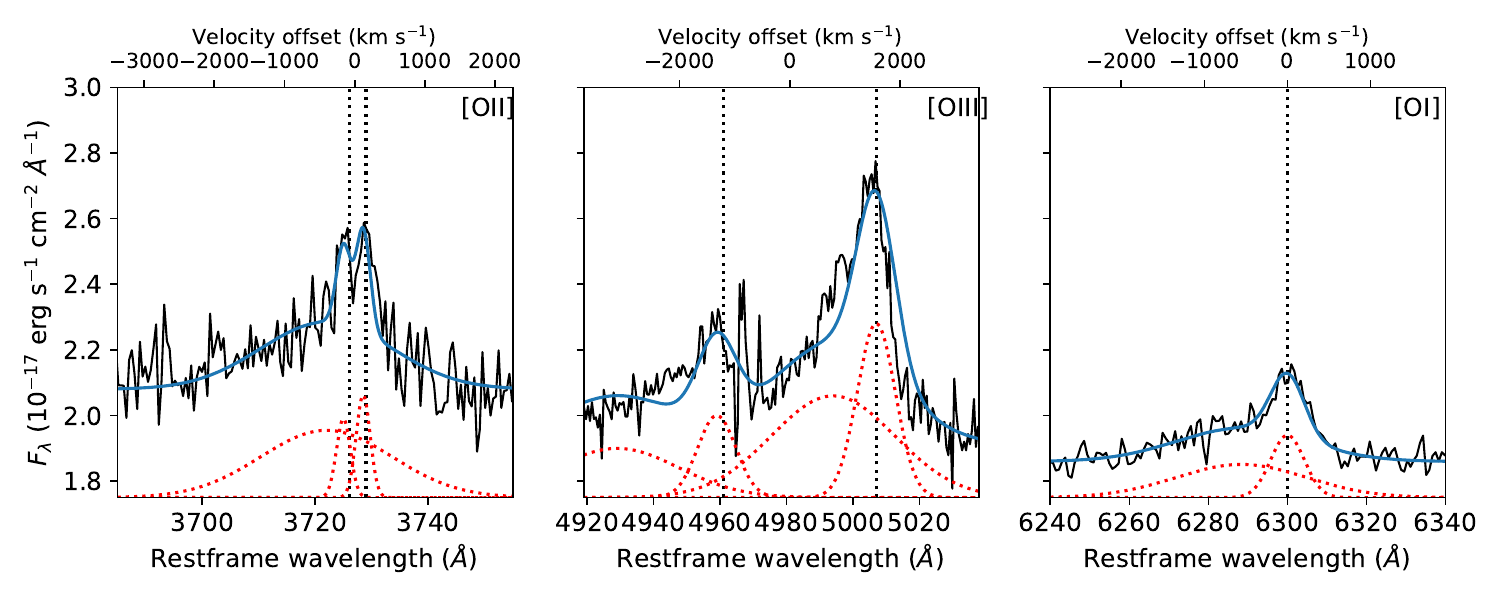}
    \caption{Zoom in of the three primary forbidden oxygen features at restframe optical wavelengths. All oxygen lines show blueshifted wings, indicative of strong winds.}    
    \label{fig:oxylines}
\end{figure*}

\clearpage
\begin{figure*}
\centering
\includegraphics[width=0.9\textwidth]{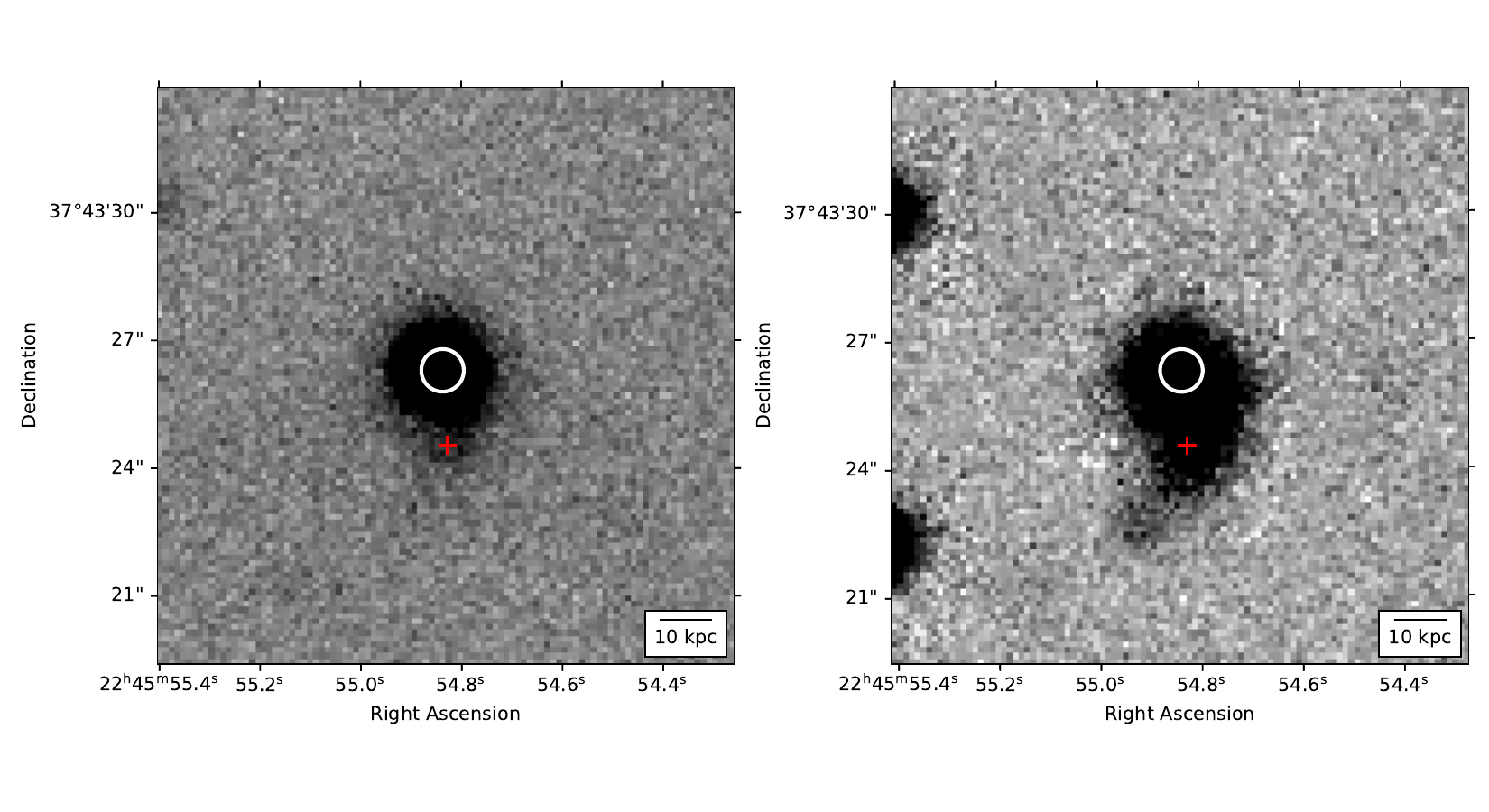}
\includegraphics[width=0.9\textwidth]{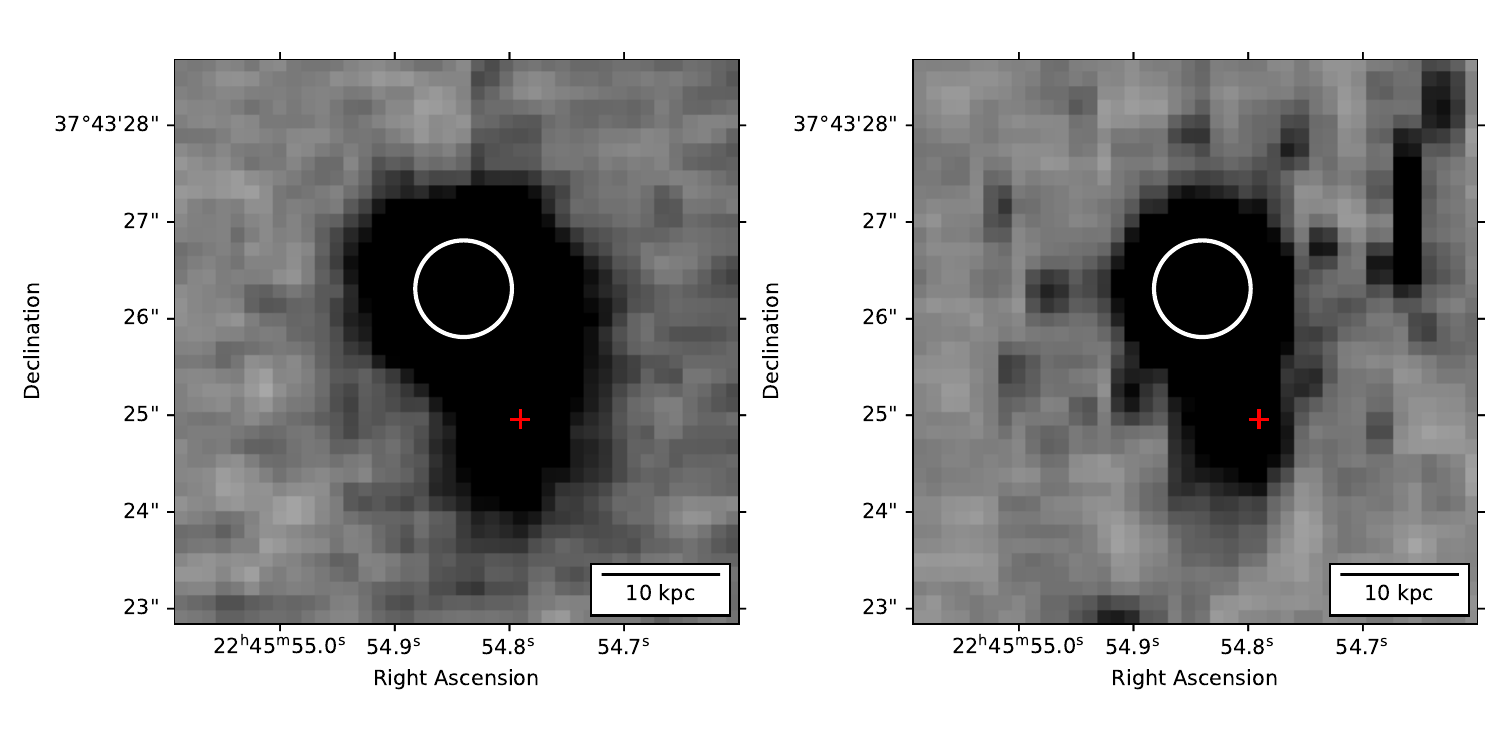}
    \caption{Imaging data for \agn. (Top): LRIS images of \agn: (left) $B$-band  and (right) $i$-band. 
    The white \as{0.5} radius circle at the position of the transient and the red cross shows the centroid of the residual from PSF-subtracted photometry, with magnitude of $i < 23.1$. 
    A scalebar indicating 10 kpc at the redshift of the AGN ($z = 2.554$) is also shown. (Bottom): KCWI imaging of \agn. The white circle indicates the position of the \as{1} radius aperture used to extract the flare spectrum. The red cross marks the position of the aperture used to extract a spectrum of the extended emission.}
    \label{fig:lrisimg}
\end{figure*}

\clearpage
\begin{figure*}
\centering
\includegraphics[width=\textwidth]{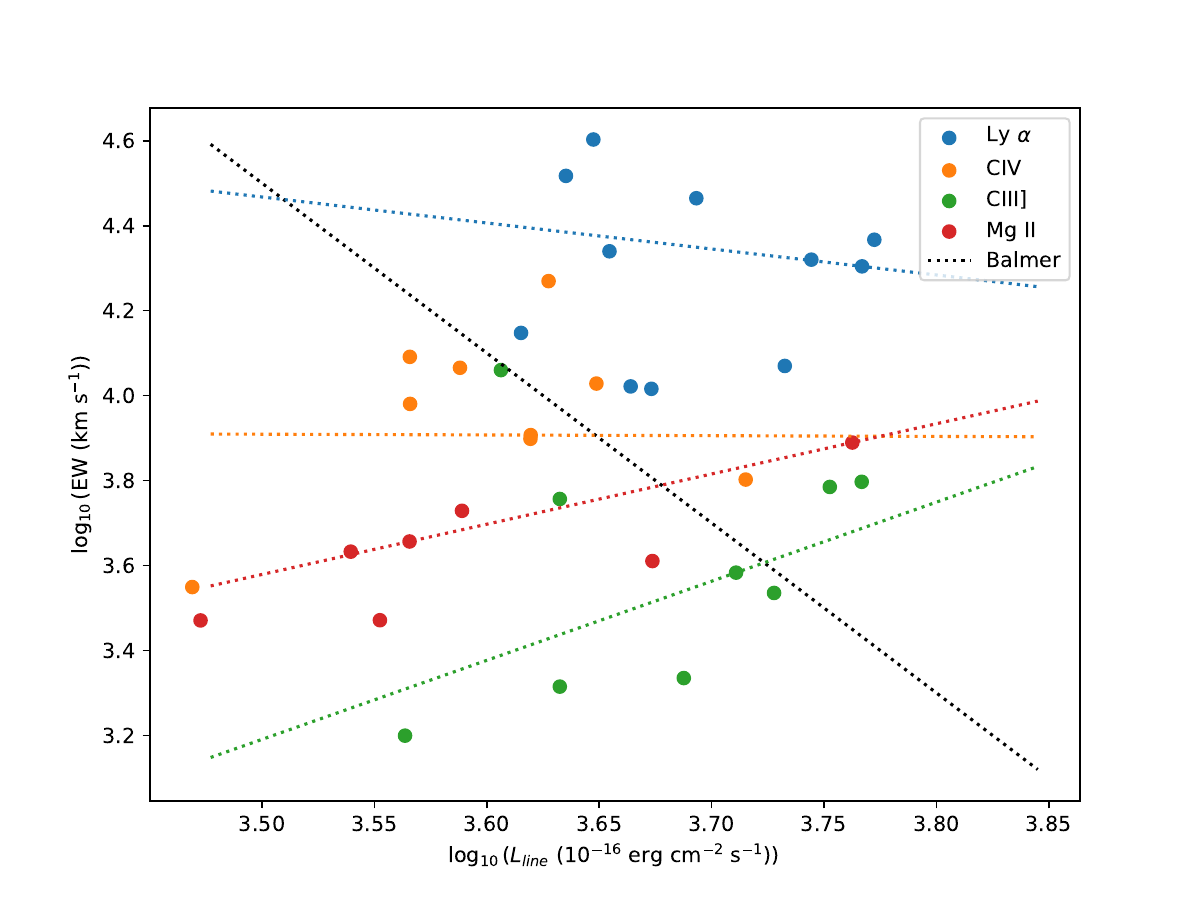}
\caption{Spectroscopic line measurements over time for \agn. Equivalent width (EW) vs. line luminosity for \Lya\ (blue), \civ\ (orange), \ciii\, (green) and \mgii\ (red) emission lines for \agn. Line width was estimated from Gaussian fits to the lines though we note that P~Cygni absorption features to both \Lya\ and \civ\ adds  uncertainty to the measurement. The dotted lines indicate the best fit linear relationship using Thiel-Sen to each line and the black dotted fit is the normal breathing relationship for Balmer lines.}
    \label{fig:breathing}
\end{figure*}

\clearpage
\begin{figure*}
 \centering
 \includegraphics[width=\textwidth]{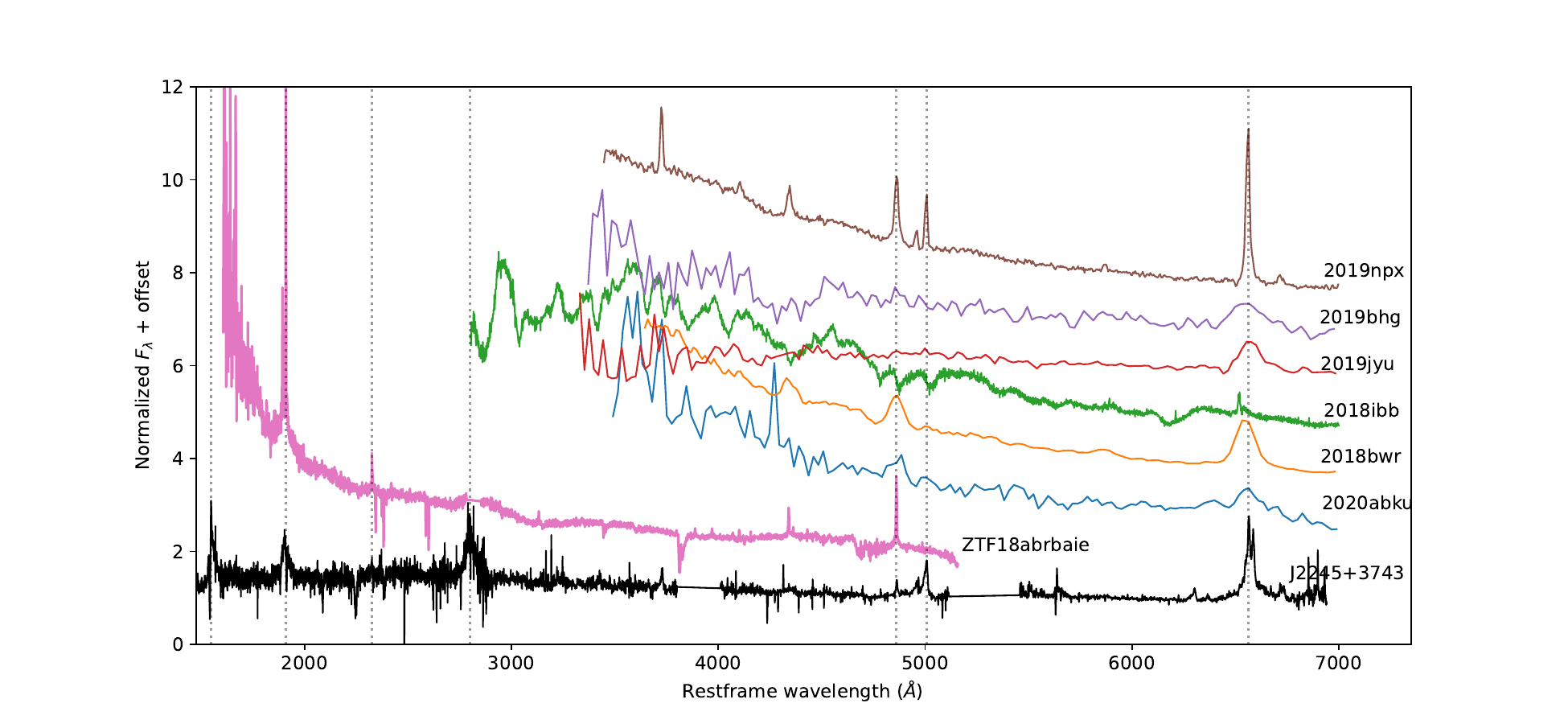}
    \caption{A comparison of the composite spectrum of \agn\ with SLSN-II spectra in the literature and the ENT AT2021lwx. \agn\ is dominated by AGN features.}
    \label{fig:snspeccomp}
\end{figure*}

\clearpage
\begin{figure*}
\centering
\includegraphics[width=\textwidth]{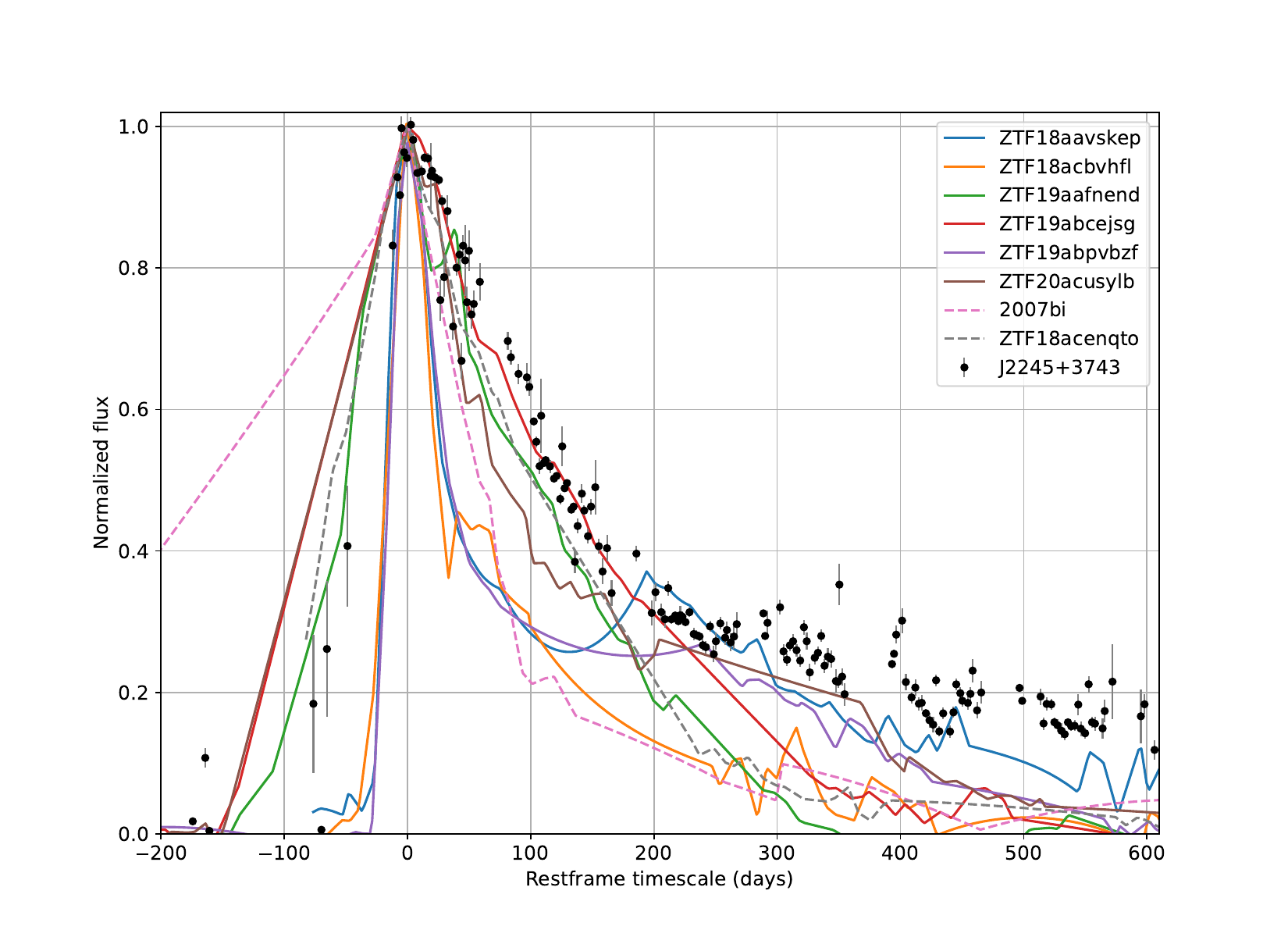}
    \caption{A comparison of the normalized light curve of \agn\ with supernovae. This shows six slow declining SLSNe (solid lines) and two candidate PISNe (dashed lines). The supernova light curves show Gaussian process fits to the photometric data with a Mat\'{e}rn-5/2 kernel. Poor light curve coverage is the cause of the structure in some of the SNe fits. Data are presented as observed values with one sigma measurement errors.}
    \label{fig:sne}
\end{figure*}



\end{document}